\documentclass[epsfig,10pt]{article}
\usepackage{graphicx}

\begin{document}

\begin{center}
{\Large\bf Seismology of the Sun : Inference of Thermal, Dynamic
 and Magnetic Field Structures of the Interior}
\end{center}

\begin{center}
{\large\bf K. M. Hiremath}
\end{center}

\begin{center}
{\em Indian Institute of Astrophysics, Bengaluru-560034, India; E-mail : hiremath@iiap.res.in}
\end{center}

\setcounter{footnote}{0}  

\begin{abstract}
Recent overwhelming evidences 
show that the sun strongly influences the Earth's climate
and environment. Moreover existence of life on this Earth
mainly depends upon the sun's energy. Hence, understanding
of physics of the sun, especially the thermal, dynamic and
magnetic field structures of its interior, is very
important. Recently, from the ground and space based observations,
it is discovered that sun oscillates near 5 min periodicity in millions of modes.
This discovery heralded a new era in solar physics and a separate
branch called {\it helioseismology} or {\it seismology of the sun} has
started. Before the advent of {\it helioseismology}, sun's
thermal structure of the interior was understood from the evolutionary solution of
stellar structure equations that mimicked the  present age, mass
and radius of the sun. Whereas solution of MHD
 equations yielded internal dynamics and magnetic field structure 
 of the sun's interior. In this presentation,
I review the thermal, dynamic and magnetic field structures
of the sun's interior as inferred by the {\it helioseismology}.
\end{abstract} 
\section{\Large\bf  Introduction }
From the dawn of civilization, sun is revered
and held as an awe inspiring celestial object in the sky. In the world, there are
many stories and poems woven around the sun god in different folklores and
the magnificent architectures are dedicated to the mighty sun.
The flora and fauna on the earth mainly depends on the
sun for their survival. Recent observational evidences
are building up that even the unpredictable climates
and rainfalls of the earth are excited and are
maintained by the sun's influence (Reid 1999;  Shine 2000; 
Hiremath and Mandi 2004 and references there in; Soon 2005; 
Hiremath 2006a, 2006c; Haigh 2007; Perry 2007; 
Feymann 2007; Tiwari and Ramesh 2007 and references there in;
Scafetta and West 2008; Hiremath 2009; Agnihotri, Dutta and Soon 2011).
Analysis of vast stretch in time
of the paleoclimatic records show  that the sun's
activity is imprinted in the global temperature and
precipitation variabilities. The imminent influence (Hiremath and Mandi 2004
and references there in; Hiremath 2009 and references there in) of the sun also
can be traced in the well documented instrumental rainfall records.

Among all the distant stars, the sun is very closest to us and is 
apparently $\sim$ 100 billion
times brighter than any other star.
Thus owing to it’s proximity and brilliance, the sun can be closely studied in
details; like it’s chemical composition, thermal and dynamic structures of the
interior, etc.
The sun is a gigantic cosmic laboratory where one can test the physical
phenomena discovered on the earth. Hence the sun is very important 
astronomical object that needs to be
studied carefully. If we understand the sun, we can understand the distant
stars and physics of whole universe.


 In the recent decade, helioseismology-a tool to probe the solar internal
structure and dynamics-has changed our conventional perception
of the internal structure and dynamics. Observational evidences show
 that sun oscillates over a wide range
of periods that range from minutes to perhaps
on time scales of centuries. Observations from the ground 
(GONG, BISON, IRIS, {\it etc.,} ) and from the space 
(like SOHO, HINODE, STEREO, etc.,) provided
solar oscillation frequency data with very high precision.

Before the advent of helioseismology, sun's internal thermal
 structure was understood from the {\em standard
solar model}. A standard solar model is built from
the evolutionary calculations of stellar structure equations
(conservation of mass, momentum and energy and supplemented
with known equation of state, knowledge of transfer of
energy by convection in the outer part of the sun,
equation of nuclear energy
generation and nature of opacity of the solar internal plasma)
that ultimately mimic the sun's present age, mass and radius
respectively. 
As for the dynamical structure, earlier  studies concentrated
on studying the rotation rate of the solar core as this parameter
is related to evolutionary history of angular momentum of the sun and its influence
on the orbit of nearby planet Mercury for testing the Einstein's
general theory of relativity. 

Similarly, models based on
the turbulent $\alpha\omega$ dynamo mechanism 
required that  radial part of the angular velocity should 
increase from the surface towards the interior in order to
reproduce the correct sunspot butterfly diagrams.
 On the other hand, results of non-linear hydrodynamical
simulations (Gilman and Miller 1981)  show that
angular velocity decreases from the surface towards the interior. These simulations
also yield internal rotational isocontours that have cylindrical geometry.
However, it will be known from the following sections that
the helioseismic inferences yield entirely different
picture of the rotational profile in the solar interior.
We will also know from the helioseismic inferences
and modeling that the thermal structure (such
as pressure, temperature, etc.,) of the solar interior
is almost similar to the structure obtained from
the {\em standard solar model}, although recent surface
chemical abundances and hence helioseismic inferences
yield the substantial disagreement.
 
Plan of this presentation is as follows. In section 2, a brief
introduction to the observational aspects of the sun
is presented. Recent discoveries of the sun's oscillations
from the ground and space are presented
in section 3. Section 4 is devoted to inferences of thermal,
dynamic and magnetic field structures of the sun's interior.
Concluding remarks are presented
in the last section. In this presentation, I mainly concentrate
on the {\em global seismology} to infer global large scale
structure of the sun's interior although one can also
infer near surface structure from the {\em local helioseismology}.

\section {A Brief Observational Introduction to the Sun}
Important physical parameters of the sun are : (i) mass-$2 \times 10^{30}$ Kg,
(ii) radius-$7 \times 10^{8}$ meter, (iii) mean density-1409 Kg$m^{-3}$,
(iv) temperature at the photosphere-5780 $^{o}K$ and, (v) the
total amount of energy radiated by the sun (i.e., luminosity)
measured at one astronomical unit (i.e., distance between
the earth and the sun)-$3.9 \times 10^{26}$ joules/sec.

When one considers the vertical cross section of the sun parallel to
it's rotation axis, based on the dynamical and physical
properties, solar interior mainly can be classified into
three distinct zones : (i) {\em the radiative core} where the
energy is mainly generated by the nuclear fusion of hydrogen
into helium and is transfered by the radiation, (ii) {\em the convection zone}
where the energy is transferred from the center to the
surface by the convection of the plasma and, (iii) {\em the photosphere}
where the energy is radiated into the space. Above the
photosphere, the sun's atmosphere consists of the
{\em chromosphere} and the {\em corona}. The temperature increases
from the layer of photosphere to million degree kelvin
in the corona.

If the sun were static in time with a constant output of energy,
the planetary environments in general and the earth's environment in
particular would have received the constant output of
energy. However, the sun's energy output
is variable due to spatial and temporal variability of the sun's large scale
magnetic field structure, dynamics and flow of mass
(both the neutral and charged particles) from the
sun. The most outstanding activity of the sun is {\em sunspots}-
cool and dark features compared to the ambient medium-on the sun's
surface that modulate sun's irradiance and the
galactic cosmic rays that enter the planetary environments.
Variation of the occurrence of number of sunspots over the surface
of the sun with an average periodicity of $\sim$ 11 years
is termed as sunspot or solar cycle. As the sunspots are
bipolar magnetic field regions with positive and negative
polarities, every 11 years polarity changes and, by
22 years, sunspots reverse back to their original polarities.
This constitutes the 22 year sun's magnetic cycle. 

The {\em flares} that are mainly associated with the sunspots (Hiremath 2006b)
 release the vast amount of energy (upto $10^{25}$ Joules) with in a
short span of time. The sun also ejects sporadically the mass ($\sim$
$10^{15}$ gm) of
plasma to the space and such an event is called {\em coronal mass ejection}.
There is also a continuous flow of wind ($\sim 10^{31}$ charged particles
per second or $6-7$ billion tons per hour) from the
sun towards the space called the {\em solar wind}.

The sun's activity varies on time scales of few minutes to months,
years to decades and perhaps more than centuries. The solar 
different activity indices vary on
the time scales of $\sim$ 27 days due to solar rotation
and $\sim$ 150 days due to the flares. Unlike the Earth,
sun rotates differentially, rotating faster near the
equator compared to the regions near the poles. The 1.3 year
periodicity is predominant in different solar activity
indices. The next prominent and ubiquitous
{\em viz.,} 11 year solar cycle periodicity is found
not only in the present day sun's activity indices but also
in the past evolutionary history derived from the
solar proxies (such as $C^{14}$ and $Be^{10}$ respectively).

Occurrence of transient events, such as either flares
or coronal mass ejections, of the solar activity
that are directed towards the earth create
havoc in the earth's atmosphere by disrupting the
global communication, reducing life time of the low-earth-orbit satellites
and, cause electric power outages.
Owing to sun's immense influence of space weather effects
on the earth's environment and climate, it is necessary
to predict (Hiremath 2008) and know in advance different physical
parameters such as amplitude and period of the future
solar cycles.

\section{Observations of the Solar Oscillations}

In a gravitationally stratified and compressible medium
such as sun and with vigorous convective activity in
the convective envelope, different types of oscillation
modes can be excited. For example, noise due to turbulence,
especially in the convective envelope, perturb (Goldreich and Kumar 1988) gradients of pressure
 and excite the sound waves or {\em p modes} that alternatively 
consist of compressions and rarefactions and, travel
with the velocity of local sound speed. Where as perturbations of the
stable radiative core excite {\em gravity or g} modes
and are due to vertical displacement of a parcel of fluid
from its normal position. In principle, these two
types of waves can be observed on the surface of the
sun through surface displacement of velocities
(by Doppler shifts of the spectrum) or intensity (temperature)
variations.
\begin{figure}
\begin{center}
\includegraphics[height=6.0cm,width=15.0cm]{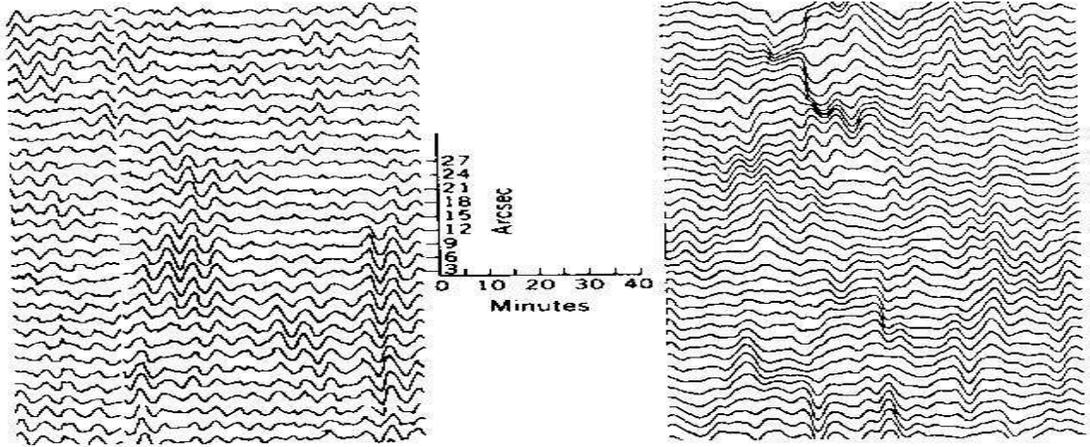}
\caption{ Observations of Doppler velocities on the solar disk,
when carried out in one spatial dimension and time, reveal the
5-min oscillations as apparent wave packets that last for
about five or six wave periods (Musman and Rust 1970).
Right figure illustrates the spatial intensity variations
 with time (Nishikawa {\em et.al} 1986).}
\end{center}
\end{figure}
Oscillations of the sun's photosphere were first discovered in the
early 1960s (Leighton 1960; Leighton {\em et. al.} 1962). 
Doppler shifts of the spectral lines formed in the atmosphere
revealed the period of oscillations of about five minutes.
Later it was discovered (Deubner 1975) from theoretical
investigations (Ulrich 1970; Leibacher and Stein 1971) that
these oscillations are due to superposition of large number
of individual modes each having its own characteristics
frequency. Surface observations indicate the velocity fluctuations
are $\sim$ 1000 m/sec. Where as amplitude of individual
modes ranges from 0.3 m/sec to 0.01 m/sec (Fig 1). Some
of the groups (Severney {\em et. al.} 1976; Kotov {\em et. al.} 1978;
 Scherrer {\em et.al.} 1979; Scherrer and Wilcox 1983) reported 
the oscillations that may represent internal gravity modes.
Observations of whole disc Doppler velocities revealed
an oscillation with a period 160.01 minute that is 
believed to be a {\em g} mode oscillation of the sun.
Thus new era in the history of solar physics
{\em viz.,} seismology of the sun or {\em helioseismology}
(as this term is initially coined by Severny {\em et. al.} 1979)
has started that is yielding rich dividends on 
physics of the sun's interior. 
\begin{figure}
\begin{center}
\includegraphics[height=8.0cm,width=8.0cm]{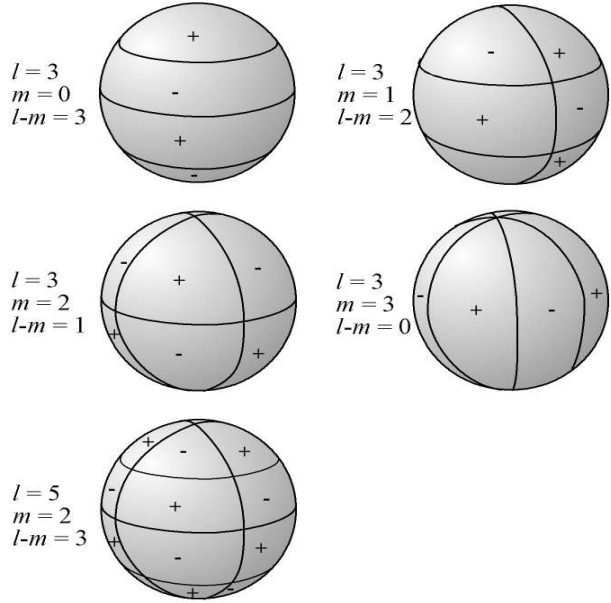}
\caption{Schematic representation of $Y_{l, m}$ on the unit sphere and its 
nodal lines. $Y_{l, m}$ is equal to $0$ along $m$ great circles 
passing through the poles, and along $l-m$ circles of equal latitude. 
The function changes sign each time it crosses one of these lines.
This figure is adopted from Wikipedia. }
\end{center}
\end{figure}

\subsection{Surface Patterns of the Global Oscillation Modes}
 In spherical coordinates, one can express velocity $V$ of oscillation
in a specific normal mode as the real part of
\begin{equation}
V(r,\vartheta,\phi,t) = V_{n}(r)Y_{l}^{m}(\vartheta,\phi)e^{-2\pi\nu t} \, ,
\end{equation}
where $Y_{l}^{m}(\vartheta,\phi)=P_{l}^{m}(cos\vartheta)e^{im\phi}$ is
spherical harmonic function of degree $l$ and azimuthal order $m$,
$\vartheta$ is co-latitude and $\phi$ is longitude of the sun.

For each pair of $l$ and $m$, there is a discrete spectrum of modes
with distinct frequency $\nu$. These differ in spatial structures
as a function of radius. $V_{n}(r)$ are the eigen functions which are
oscillatory in space and typically possess $n$ zeros or nodes
in the radial direction. The degree $l$ can be understood as the
number of complete circles on the surface of a star where the
normal component of velocity is always zero.

The angular degree $l$ is related to the total horizontal component
of the wave number, $K_{h}$ as
\begin{equation}
K_{h}=\frac{[l(l+1)]^{1/2}}{R_{\odot}} \, ,
\end{equation}
where $R_{\odot}$ is the radius of the star.
Thus high degree $l$ modes sense the near surface
structure and very low degree $l$ modes sense
near center of the sun. Thus with different degree
modes observed on the surface, one can infer
the internal structure of sun.
The azimuthal order m, characterizes the orientation of the mode
relative to the axis of the spherical coordinate system and has 
$(2l+1)$ values ({\em i.e.,} $-l, ...., l)$. The zero velocity lines
are like lines of latitude when $m=0$ and are like lines of
longitude when $m=\pm l$. When $m=0$, the spherical
harmonic is called as {\em zonal harmonics} and
the harmonic with $m=l$ is called {\em sectoral harmonics}. {\em Tesseral
harmonics} is obtained when $m \ne l$ (see Fig 2).
\subsection{Amplitudes, Frequencies and Line widths of 5 min Oscillations}
There are two ways, viz., {\em unimaged} and {\em imaged} methods,  of measuring the amplitudes
and frequencies of 5 min oscillations (Hill, Deubner and Isaak 1991). In case of {\em unimaged} method, 
oscillations are due to low-$l$ modes that are
obtained from the integrated whole disk measurements.
These modes have longer life times ($\sim$ of months)
and hence their frequencies can be accurately measured.
\begin{figure}
\begin{center}
\includegraphics[height=8.0cm,width=10.0cm]{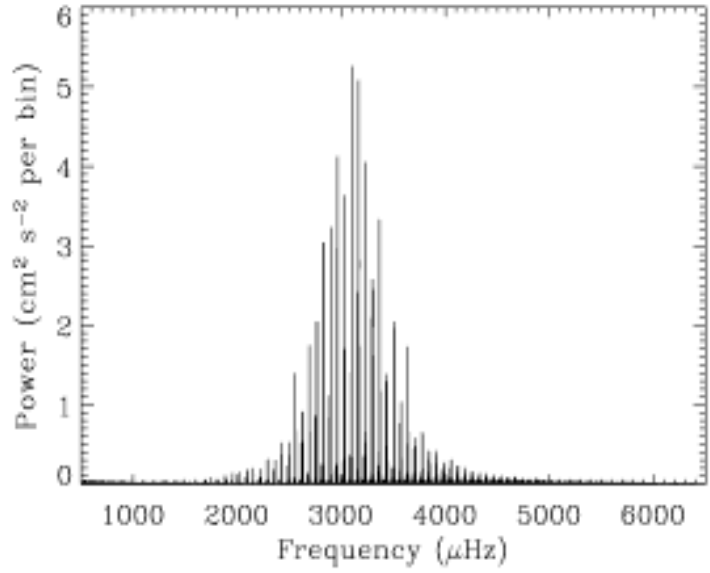}
\caption{ Power spectrum of integrated low degree $p$ modes (Deubner and Gough 1984).}
\end{center}
\end{figure}

Fig 3 illustrates an example of low-$l$ power
spectrum obtained from Birmingham integrated whole
disk measurements.
One can notice from Fig 3 that, significant power lies between the 1-5 mHz.
From the spherical Fourier analysis, it is concluded
that these observed oscillations are identified as low 
degree ($l=0, 1, 2$ and $3$) and low radial order ($0 \le n \le 10$) 
acoustic modes that  penetrate deeply the solar interior. 
\begin{figure}
\begin{center}
\leftmargin -5cm
{
\includegraphics[height=8.0cm,width=8.0cm]{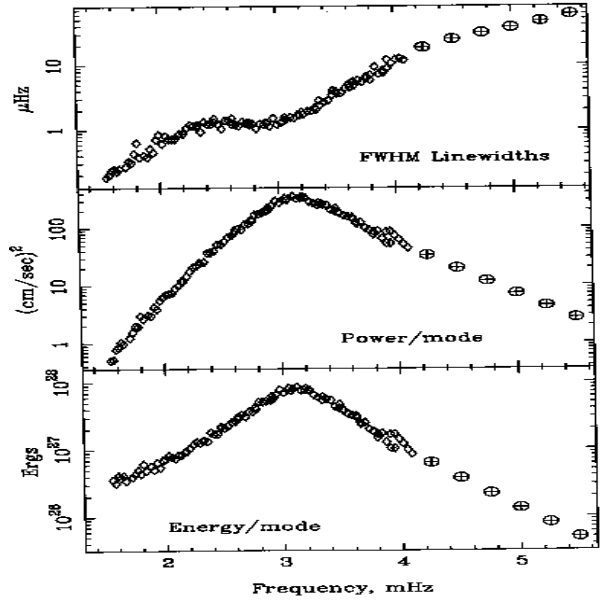}
}
\caption{ Line width (top), power per mode (middle)
and total energy per mode (bottom) as a function
of frequency $p$ modes for the modes $l=20$ are illustrated (Libbrecht 1988).}
\end{center}
\end{figure}

As for the measured amplitudes of oscillations in Doppler velocity,
it depends upon the choice of the spectral lines (that occur 
at different heights )
used for sensing the oscillations. For example,
in potassium and sodium lines, maximum amplitudes from the power spectra  
are estimated to be $\sim$ 15 and 25 $cm^{-1}$ respectively.  
In case of irradiance, amplitude of oscillations
is estimated to be $\sim$ 2-3 x $10^{-6}$.
One can get the tables of amplitudes of oscillations
from the observations of Doppler velocity measurements (Grec {\em et. al.} 1983;
Palle {\em et.al.} 1989; and Anguera Guba {\em et. al.} 1990).
 
In case of {\em imaged} method, amplitudes and frequencies of the 
solar oscillations are computed from the spatially resolved disk
of the sun. With this method, amplitudes and frequencies
of many individual modes can be computed. For example, from the
information of measured amplitudes, one can get the information of  power
and energy per mode. For the degree $l=20$,
power and energy per mode of oscillations as a function of frequency
are illustrated in Fig 4.

Another useful physical parameter measured from the
{\em nonimaged} and {\em imaged} observations
is the line width of different oscillation modes that provide   
vital information of excitation and damping
mechanism of the solar oscillations. Measurements
of line widths suggests that line width increases
with both the frequency ($\nu$) and degree ($l$)
of the oscillations. For example, for $l=20$, in Fig 4, 
FWHM (Full width at half maximum) of the
line widths as a function of frequency is illustrated.  
When one compares the observations of line width
as a function of frequency and theoretical estimates
(Goldreich and Kumar 1988; Kumar {\em et. al}. 1988;
Kumar and Goldreich 1989; Dalsgaard {\em et. al.} 1989),
it is found that growth of the individual modes probably
is constrained by the perturbations of the oscillations
rather than non-linear interaction among the individual
modes.
\begin{figure}
\begin{center}
\includegraphics[height=6.0cm,width=7.0cm]{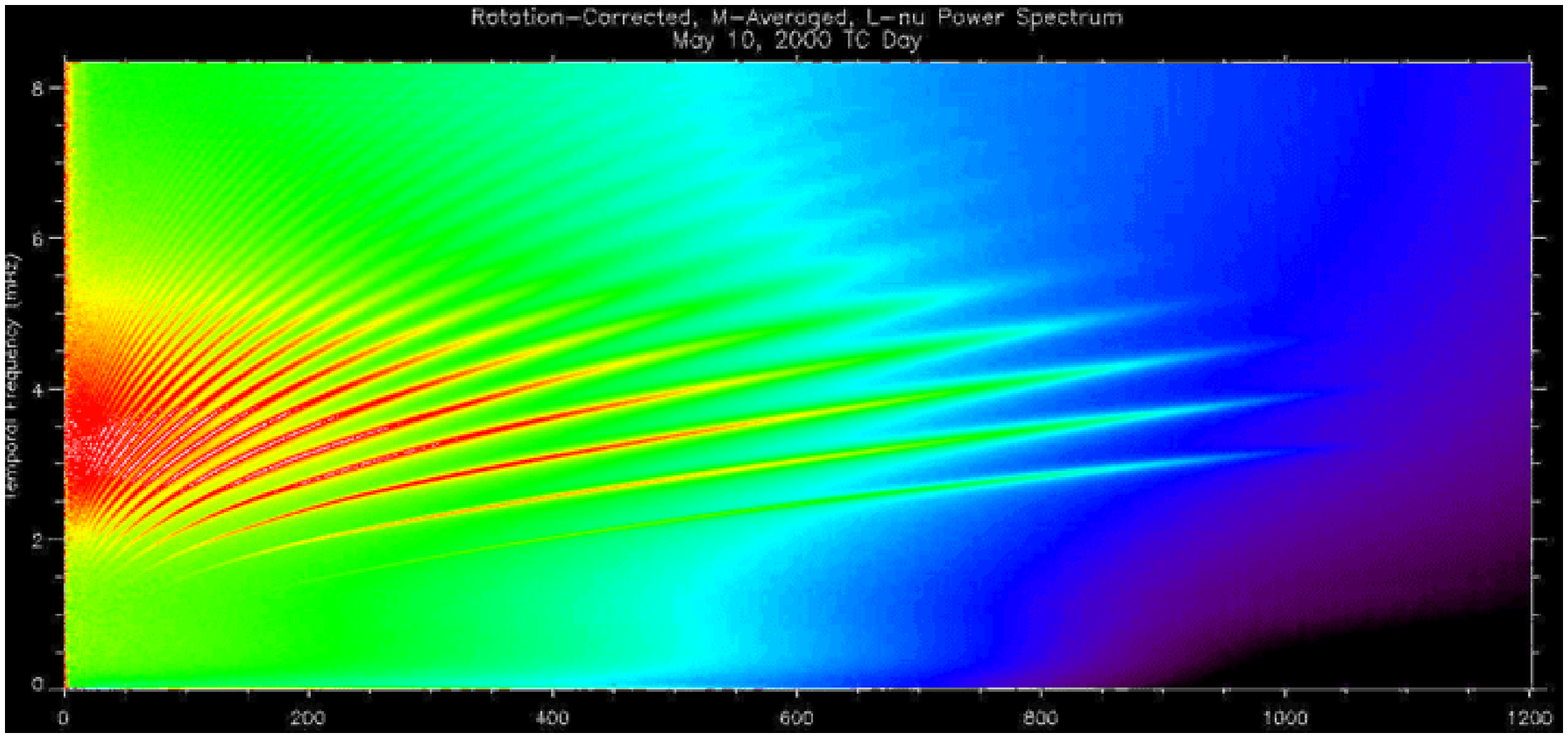}
\includegraphics[height=6.0cm,width=7.0cm]{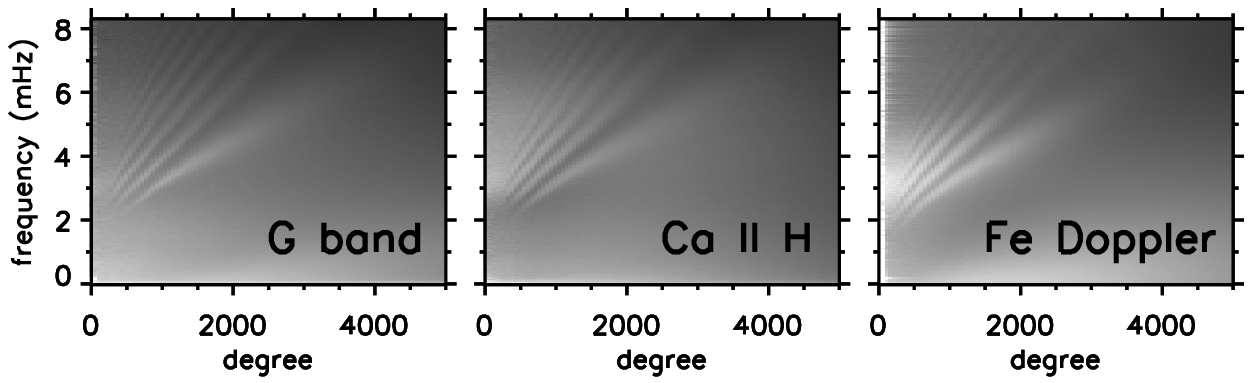}
\caption{ Typical $l$-$\nu$ diagnostic diagram of the $p$ modes: (i) on the
left from one day of GONG observations (image courtesy NSO/GONG, Howe) and,
on the right side one day Hinode observations (image courtesy Hinode/SOT, Sekii).} 
\end{center}
\end{figure}

\subsection{Observed l-$\nu$ν Diagnostic Power Spectrum of the Oscillations}
For the {\em imaged} observations, continuous
and long series of oscillation data set is subjected
to spherical harmonic Fourier analysis, averaged
over different $m$ azimuthal modes that ensures
the removal of effect of sun's rotation and l-$\nu$ diagnostic ν
power spectrum of the solar oscillations is obtained.
In Fig 5, such a l-$\nu$νdiagnostic power spectrum
is illustrated. In this figure, one can notice that
maximum power is concentrated near a frequency
of 3 mHz that corresponds to a period of 5 minutes.
Another notable interesting characteristic property
of this l-$\nu$ν diagnostic diagram is that
concentration of power is not distributed randomly,
 rather it is systematically concentrated
in the narrow curved ridges indicating that observed
oscillations are due to internal standing waves
(confined within the resonant cavity) whose
amplitudes vary from central core to near the surface. 

\subsection{Frequency Splittings due to Rotation and Magnetic Field}
In the previous section, solar power spectrum
is presented by removing the effect of rotation. 
Presence of rotation, Coriolis force and magnetic field 
of the sun affect on the dynamics of the oscillations
and hence on the frequencies. For example, if the
sun is not rotating, non-magnetic and spherically symmetric, 
then the modes with same $n$ and $l$,
oscillation frequencies would be degenerate
in azimuthal order $m$. However, sun is rotating
and magnetic resulting in lifting the degeneracy
that makes the frequencies depend upon the
azimuthal order $m$. In case of sun, compared to
magnitudes of rotation and magnetic field structures, effects
due to Coriolis and centrifugal forces are negligible. 
From the observed data, rotational frequency splittings are computed as
$\delta \nu_{nlm} = \nu_{nlm} - \bar \nu_{nl}$, where
$\bar \nu_{nl}$ is average frequency for $m=0$.
In Fig 6, typical power spectrum of solar oscillations, 
for different $m$ with $l=20$, obtained from the
Big Bear Solar Observatory (Libbrecht 1989) is presented.
Clearly one can notice the frequency splittings due to
rotation in this illustration.  
\begin{figure}
\begin{center}
\includegraphics[height=8.0cm,width=12.0cm]{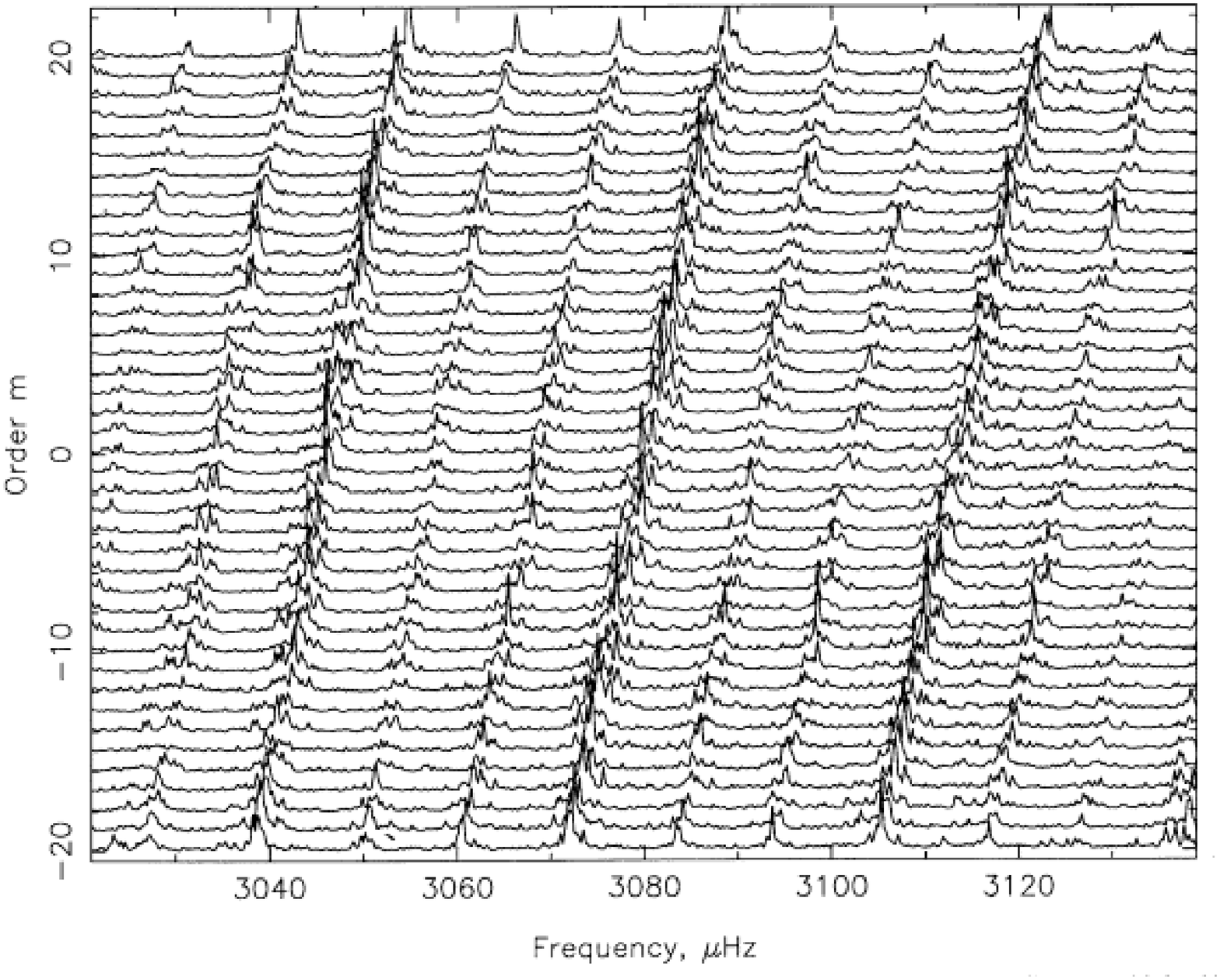}
\caption{For l=20 and different m, observed frequency splittings due 
to rotation of the sun (Libbrecht 1989). 
}
\end{center}
\end{figure}

Traditionally frequency splittings
are related to frequency $\nu_{nlm}$ of a mode as
follows
\begin{equation}
\nu_{nlm} = \nu_{nl} + \sum_{i=1}^{i_{max}}a_{i}(nl)P_{i}^{nl}(m) \, , 
\end{equation}
where $a_{i}$ are the splitting coefficients and $P$ are the
polynomials related to Clebsch-Gordon coefficients. 
Odd degree splitting coefficients $a$ are due to  
rotational effects and the combined effects (of perturbations 
due to magnetic field structure, structural asphericities
and second order effect due to rotation) contribute to the
even degree splitting coefficients. 
\section{Helioseismic Inferences of Thermal, Dynamic and Magnetic field Structures}
Since the discovery of sunspots by Galileo, scientists' quest is to
understand the physics of the sun's interior.
From the helioseismic observations, one can
precisely estimate the frequencies and their splittings of different
modes of oscillations.
For example, by knowing difference between computed theoretical 
and observationally estimated frequencies, one can infer
thermal structure such as internal sound speed and density of the
solar interior. Where as the frequency splittings 
are used for the inference
of internal dynamics (such as steady and time dependent
parts of rotation and angular momentum ) and magnetic
field structures respectively.
\subsection{Inference of Thermal Structure by Comparing the
Observed and Computed Frequencies}
Thermal structure of the solar interior
can be understood in the following two ways.
By using information of
macro and micro physics of the solar interior, compute theoretical 
frequencies (Dalsgaard 2003; Unno {\em et. al}
1989)  and match with the observed frequencies.
Macrophysics involves standard hydrodynamic equations
that are to be linearized and the perturbed
variables are assumed to vary with $e^{i \omega t}$,
where $\omega$ is angular frequency. 
Where as  micro physics requires other additional
details of the solar interior, {\em viz.}, equation of state,
opacity and rate of nuclear energy generation. 
Using macro and micro physics, with additional
knowledge of physics of the convection, solar
internal structure such as pressure, temperature, density, etc.,
are computed.

As the observed amplitudes
of oscillations are very small, equation
of energy can be neglected and, linearized
adiabatic equations with appropriate boundary
conditions are used to compute the theoretical
frequencies and are matched (Dalsgaard 2003; Unno {\em et. al}
1989) with the observed frequencies. Although most of the observed frequencies
perfectly match very well for the low and intermediate
degree modes, computed frequencies do not match
very well with the high degree modes. Possibly
this suggests our poor knowledge of physics of
near surface effects.
\subsubsection{Inference of Thermal Structure: Primary Inversions}
In another method, observed frequencies
and model of the solar structure are used to
invert the thermal structure (for example,
sound speed and density) of the solar
interior in the following way. 
For the case of adiabatic
oscillations, with additional constraint of
conservation of mass, generalized form of equation (Lynden-Bell and
Ostriker 1967; Gough and Taylor 1984; Unno {\em et. al.} 1989; Dalsgaard 2003)
of oscillation that takes into account the effects of velocity
flows $\vec{v}$ and magnetic field structure $\vec{B}$ is given as follows
\begin{equation}
\mathcal L (\vec {\xi}) - \rho \omega^{2} \vec {\xi} + \nabla  \delta p 
+\rho[\omega \mathcal M (\vec {\xi}) 
+ \mathcal N (\vec {\xi}) + \mathcal B (\vec {\xi})] = 0 \, ,
\end{equation}
\noindent where 
\begin{equation}
 \mathcal L (\vec{\xi}) = \nabla (c^{2} \rho \nabla . \vec \xi 
+ \nabla P. \vec \xi) - g \nabla.(\rho \vec \xi)
- G \rho \nabla (\int_{V} {\nabla.(\rho \vec \xi)dV \over |\vec r - \vec r^{'}|})  \, ,
\end{equation}  
\begin{equation}
\mathcal M (\vec {\xi}) = 2 i [\vec{\Omega}_{0} \times \vec {\xi} + (\vec{v}\cdot\nabla)\vec {\xi}] \, ,
\end{equation}
\begin{equation}
\mathcal N (\vec {\xi}) = (\vec{v}\cdot\nabla)^{2}\vec {\xi} 
- 2 \vec{\Omega}_{0} \times [(\vec {\xi}\cdot\nabla)\vec{v} - (\vec{v}\cdot\nabla)\vec {\xi}]
- (\vec {\xi}\cdot\nabla)(\vec{v}\cdot\nabla)\vec{v} \, ,
\end{equation}
\begin{equation}
\mathcal B (\vec {\xi}) = (4 \pi \rho)^{-1}\big[[\rho^{-1}(\vec {\xi}\cdot\nabla)\rho
+ \nabla\cdot\vec {\xi}]\vec{B} \times (\nabla \times \vec{B})
- [(\nabla \times \vec{B}^{'})\times\vec{B} 
+ [(\nabla \times \vec{B})\times\vec{B}^{'}]\big] \, ,
\end{equation}
\noindent where $\mathcal L$ is differential operator, $\vec \xi$ is eigen function of
oscillations, $c$ is speed of sound, $\rho$ is density, $P$ is pressure,
$\mathcal M (\vec {\xi})$ and $\mathcal N (\vec {\xi})$ are effects due
to velocity flows, $\mathcal B (\vec {\xi})$ is effect due to magnetic field structure
and other symbols have usual meanings. 
Although effects due to flows and magnetic field
structure in the solar interior can not be neglected, for
simplicity, these terms are neglected and with special boundary conditions such that density
and pressure perturbations completely vanish at the surface,
equation (4) leads to the form $\mathcal L (\vec {\xi}) = \rho \omega^{2} \vec {\xi}$. 
This equation is an eigen value problem and  is also Hermitian.
Hence, variational principle (Chandrasekhar 1964; see also Unno {\em et. al.} 1989)
can be used to linearize the equation and
frequency difference $\delta \nu_{n,l}$ between the sun and
standard solar structure model can be represented (Basu 2010 and references there in)
as follows 
\begin{equation}
{\delta \nu_{n,l} \over \nu_{n,l}} = \int_{0}^{R} K_{c^{2},\rho}^{nl}(r) 
{\delta c^{2} \over c^{2}}(r) dr + \int_{0}^{R} K_{\rho, c^{2}}^{nl}(r)
{\delta \rho \over \rho}(r) dr + {F(\nu_{n,l}) \over E_{nl}} \, ,
\end{equation} 
\noindent where $K_{c^{2},\rho}^{nl}(r)$ and $K_{\rho, c^{2}}^{nl}(r)$
are kernels that involve eigen functions
of the oscillations and the known solar structure. 
Other terms $\delta c^{2} \over c^{2}$ and $\delta \rho \over \rho$
 are the differences of sound speed
and density between the sun and the reference solar structure model.
The last term (see for details Basu 2010) is a correction due to surface effects and
does not arise due to linearization. 
Hence, from the observed oscillation frequencies
and with the known reference solar structure model,
one can invert radial profiles of $\delta c^{2} \over c^{2}$ 
and $\delta \rho \over \rho$ in the solar interior.
There are many inversion techniques (Gough and Thompson 1991;
Unno {\em et. al.} 1989; Dalsgaard 2002 and references there in) to infer these
profiles and one such inversion that yields the differences 
of sound speed and density (between the sun and the solar structure model) 
is presented (Basu {\em et. al.} 2009) in Fig 7. 
\begin{figure}
\begin{center}
\includegraphics[height=6.0cm,width=7.0cm]{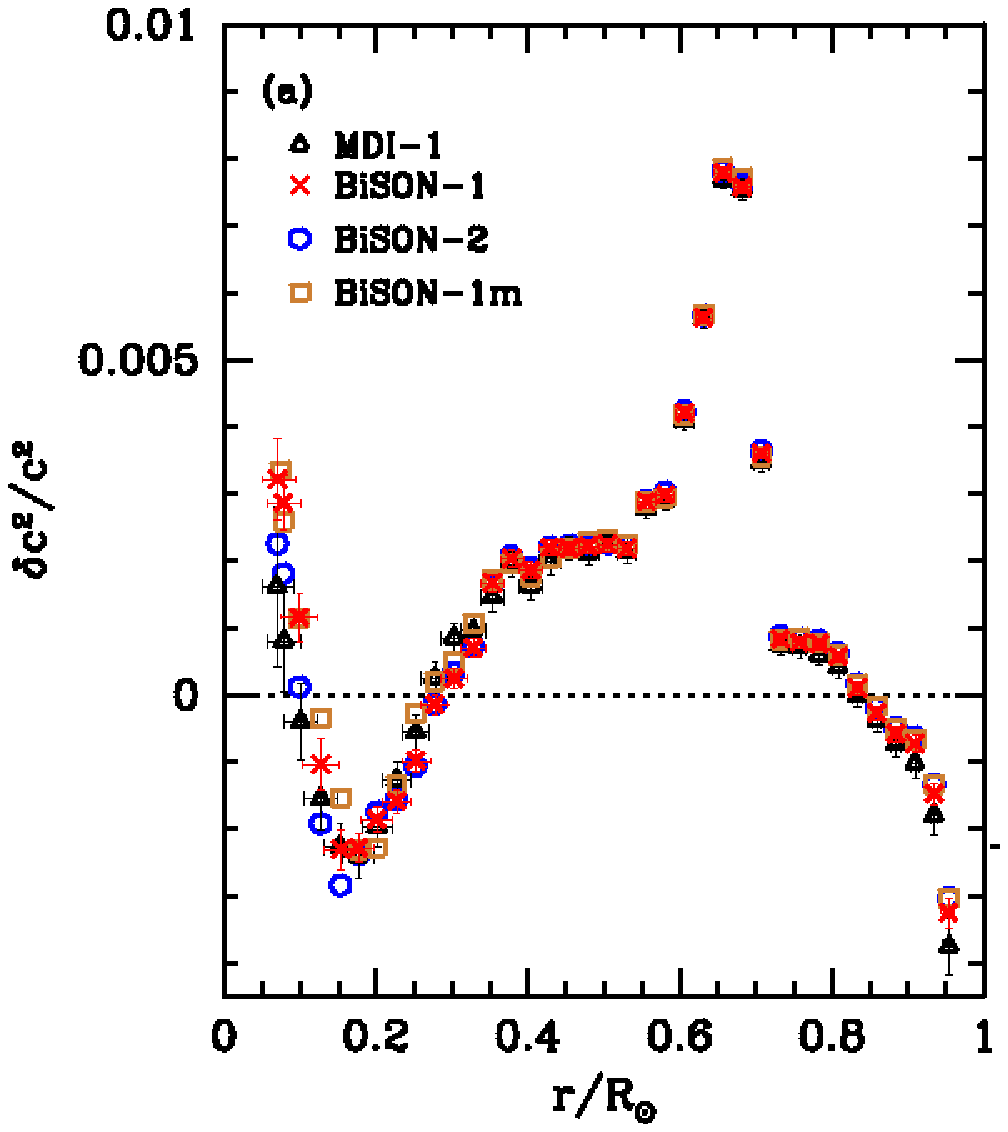}
\includegraphics[height=6.0cm,width=7.0cm]{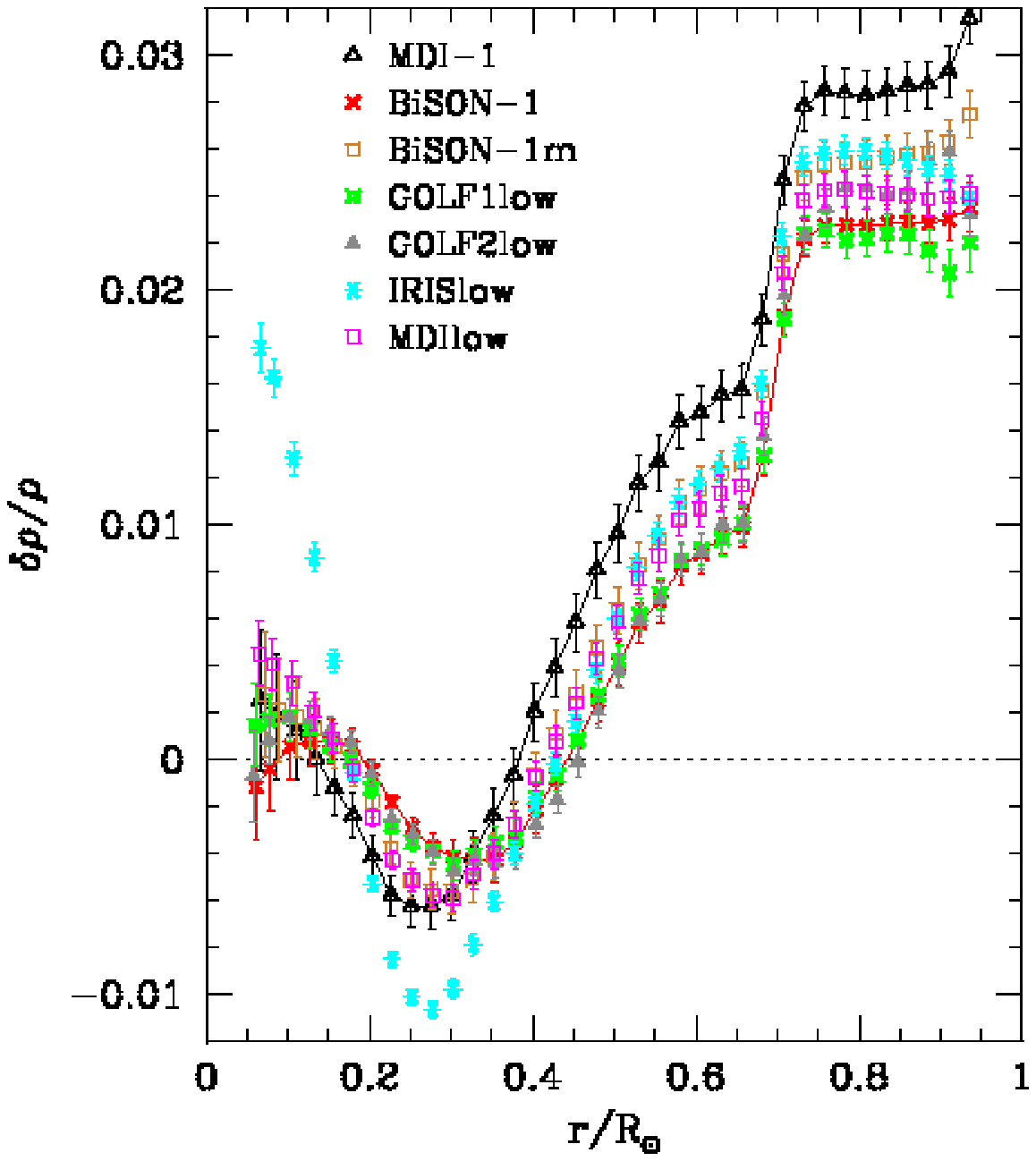}
\caption{Inference of thermal structure: (i) Left figure illustrates
the sound speed difference between the sun and the model and, (ii)
right figure illustrates the density difference between the sun
and the model. Image courtesy, Basu.}
\end{center}
\end{figure}
It is interesting to note that
most of the models (Basu 2010 and references there in)
almost match with the reference solar structure models, yet
there are unexplained statistically significant sound speed
differences mainly near base of the convection zone 
and in the radiative core. Similarly, density profile of solar
structure models developed so far do not match very well 
with the density profile of the real sun. All of the solar structure models 
require the surface chemical heavy elemental abundances
$Z$ (or ratio $Z/X$, where $X$ is the hydrogen abundance) that indirectly
contribute to the equation of state and the opacity
and hence, ultimately affect the computed frequencies,
modeled sound speed and density structures.
Most of these models (Basu 2010 and references there in)
used the heavy elemental abundance ratio
($Z/X$=0.0229) by Grevesse and Noels (1993).
However, recent determination (Asplund {\em et.al.} 2000, 2004;
Allende Prieto {\em et.al} 2001, 2002) lowered
this ratio ($Z/X$ = 0.0166) leading to further
deterioration of match between sound speed
of the sun and different models. Hence, this conundrum
 of unexplained differences of thermal structures of
sound speed and density is remained to be explained.
\subsubsection{Inference of Thermal Structure: Secondary Inversions}
By knowing the reference solar structure model,
in the previous section, observed frequencies are
used to compute the thermal structure differences of sound speed
and the density. From these differences and knowing the
sound speed and density structures as determined
by the reference models, one can obtain sound speed
and density profiles of the sun. As the sound
speed and the density structures are thermodynamic
quantities, in the following, using stellar structure
equations (with additional information of equation
of state and the opacity structures), one can uniquely
obtain the thermal structure of the solar interior.
There are two advantages of this solar seismic model, as
in the standard solar models,  (i) no assumption of
history of the sun and, (ii) no need to have adhock
computation of the convective flux. In addition,
chemical compositions such as hydrogen and helium
abundances are obtained as part of solution. Thus,
in essence, this seismic model yields a
snap shot model of the present day sun.

For example, if equation of state of the solar plasma
is available, inverted sound speed can be constrained to yield
the ratio $T/\mu$ (where $T$ is temperature and $\mu$
is the mean atomic weight).  
Following are the four set of stellar structure differential
equations that need to be solved by imposing sound 
speed obtained by the primary inversions
\begin{equation}
{dM_{r} \over dr} = 4 \pi r^{2} \rho \, ,
\end{equation}

\begin{equation}
{d P \over dr} = - {G M_{r} \rho \over r^{2}} \, ,
\end{equation}

\begin{equation}
{dL_{r} \over dr} = 4 \pi r^{2} \rho \epsilon \, ,
\end{equation}

$$
{d T \over dr} = -\frac{3}{4ac} \frac{\kappa \rho}{T^{3}}\frac{L_{r}}{ 4 \pi r^{2}} \,\,\,\,\,\,
\,\,     if \, \, radiative \, ,
$$
\begin{equation}
   = ({d T \over dr})_{conv} \, \, \, \, \, \, if \, \, convective  \, ,
\end{equation}
\noindent where the radial variables $M_{r}$, $P$, $L_{r}$
and $T$ are the mass, pressure, luminosity and temperature
respectively. Other variables $\epsilon$, $\kappa$
and $({d T \over dr})_{conv}$ are the rate of nuclear
energy generation, opacity of matter and knowledge
of convective energy transport in the outer 30 $\%$
of the sun. Further axillary equations such as
equation of state, equations of opacity and rate of 
nuclear energy generation are required. 
It is assumed that sun is in mechanical and
thermal equilibrium such that whatever energy generated
in the deep radiative core must be transported
to the surface. However this latter condition can not
be guaranteed.
 
As the sound speed is thermodynamically determined
quantity, it is a function of other three variables, {\em viz.,}
pressure, temperature and the chemical composition.
As for the chemical composition, hydrogen $X$ and helium 
$Y$ are separately considered and all other elements are treated as single
entity and is called as heavy elemental abundance $Z$.
\begin{figure}
\begin{center}
{
\includegraphics[height=5.0cm,width=5.0cm]{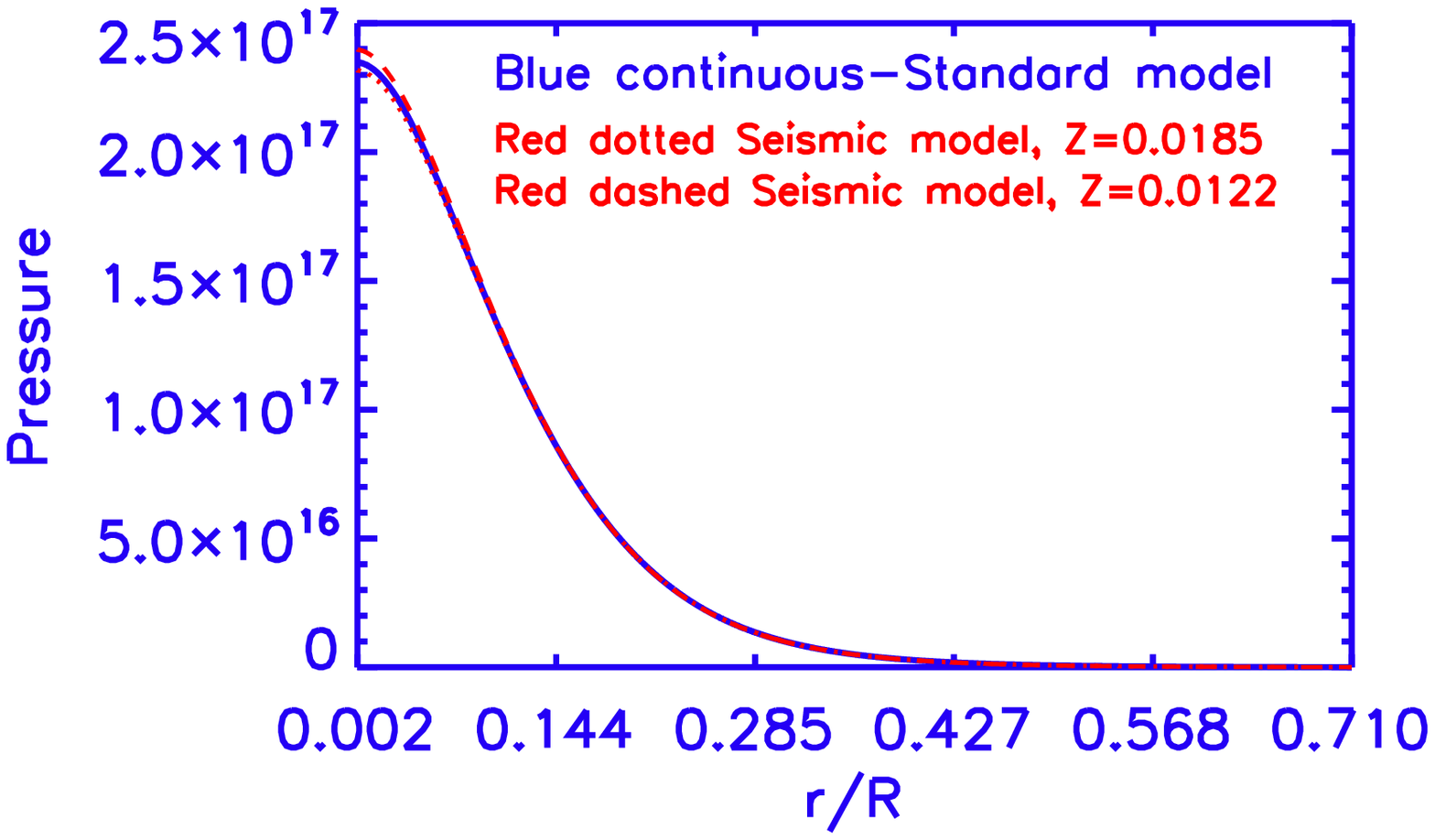}
\includegraphics[height=5.0cm,width=5.0cm]{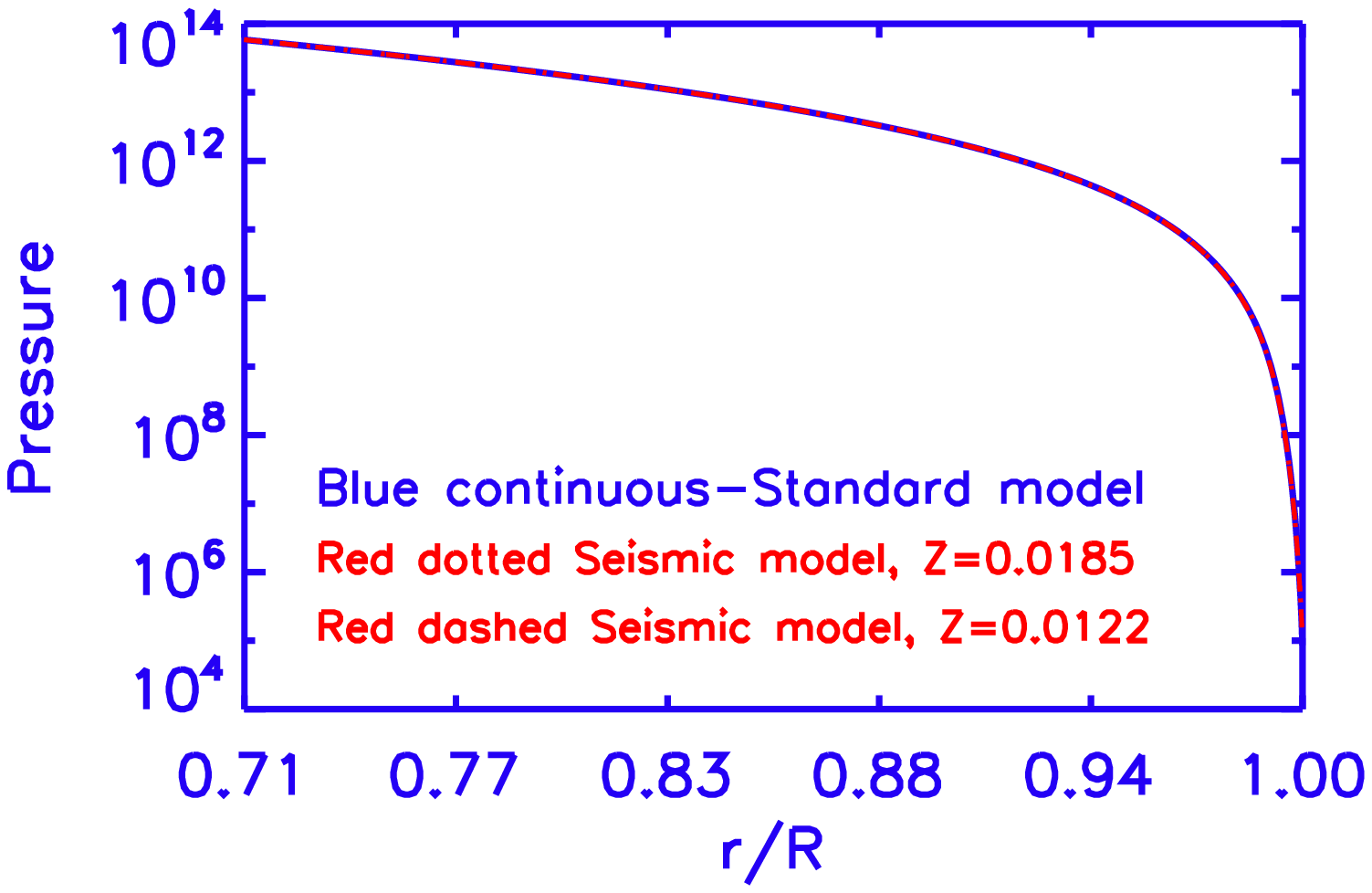}
\includegraphics[height=5.0cm,width=5.0cm]{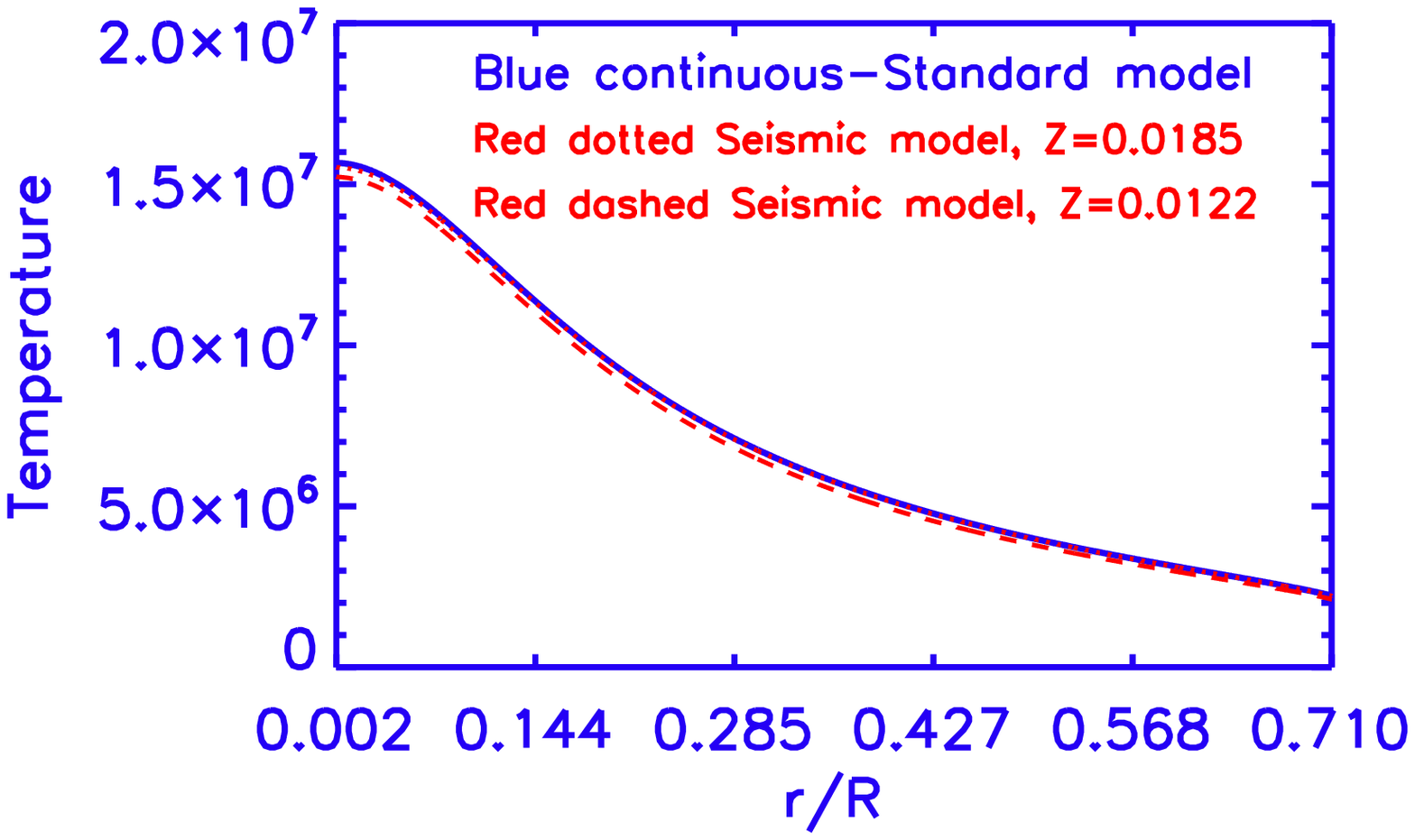}
\includegraphics[height=5.0cm,width=5.0cm]{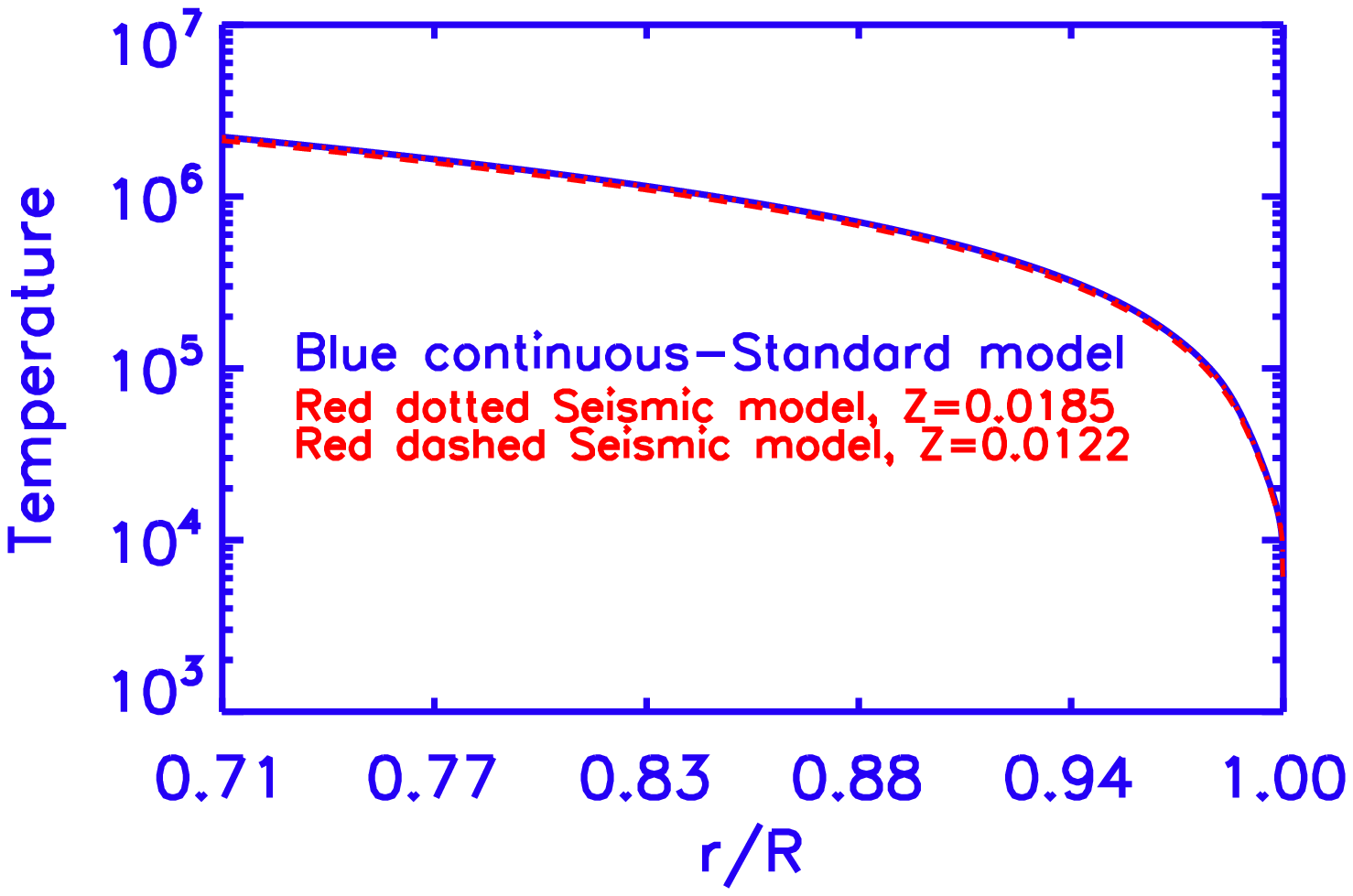}
\includegraphics[height=5.0cm,width=5.0cm]{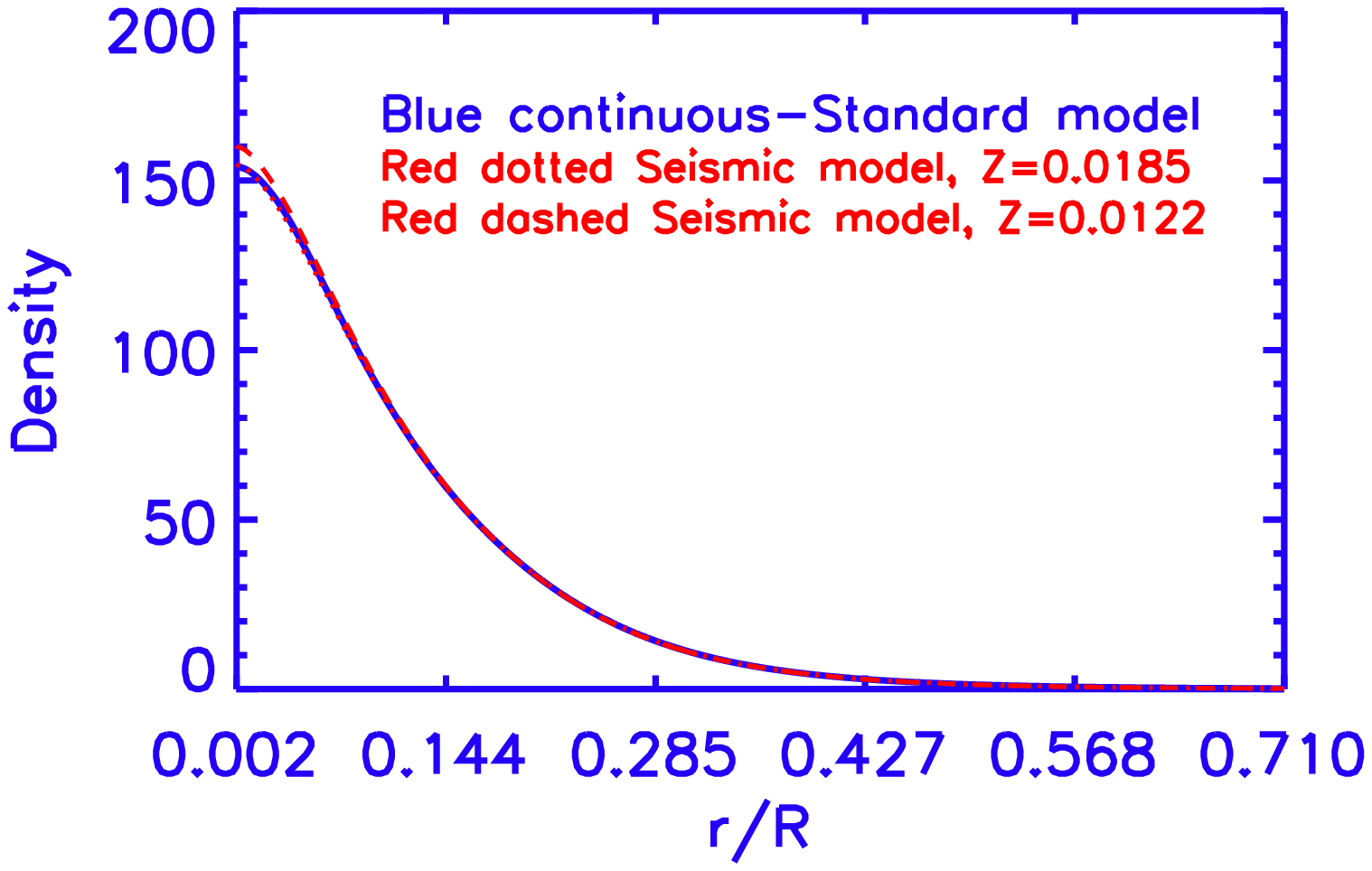}
\includegraphics[height=5.0cm,width=5.0cm]{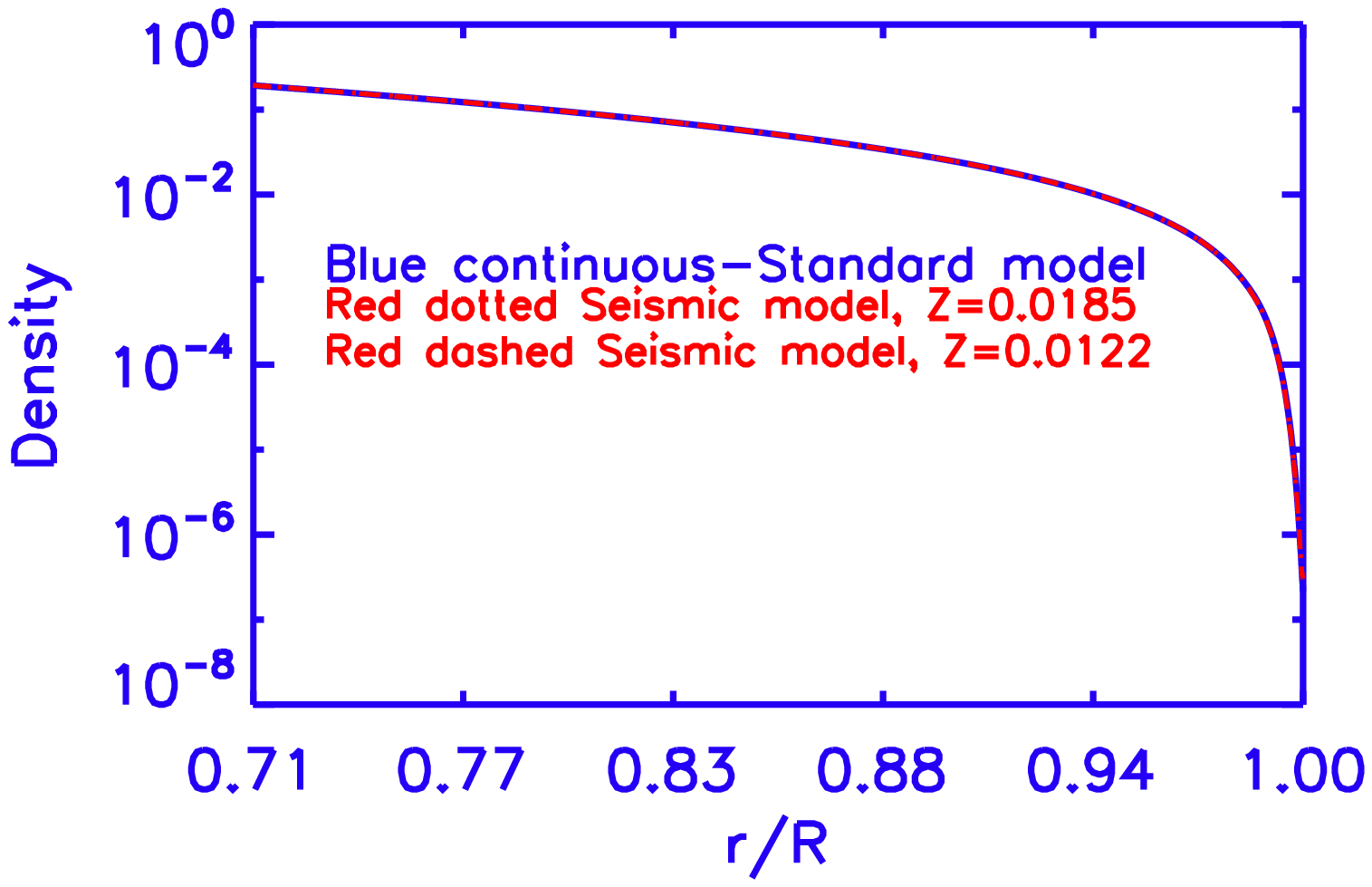}
\includegraphics[height=6.0cm,width=7.0cm]{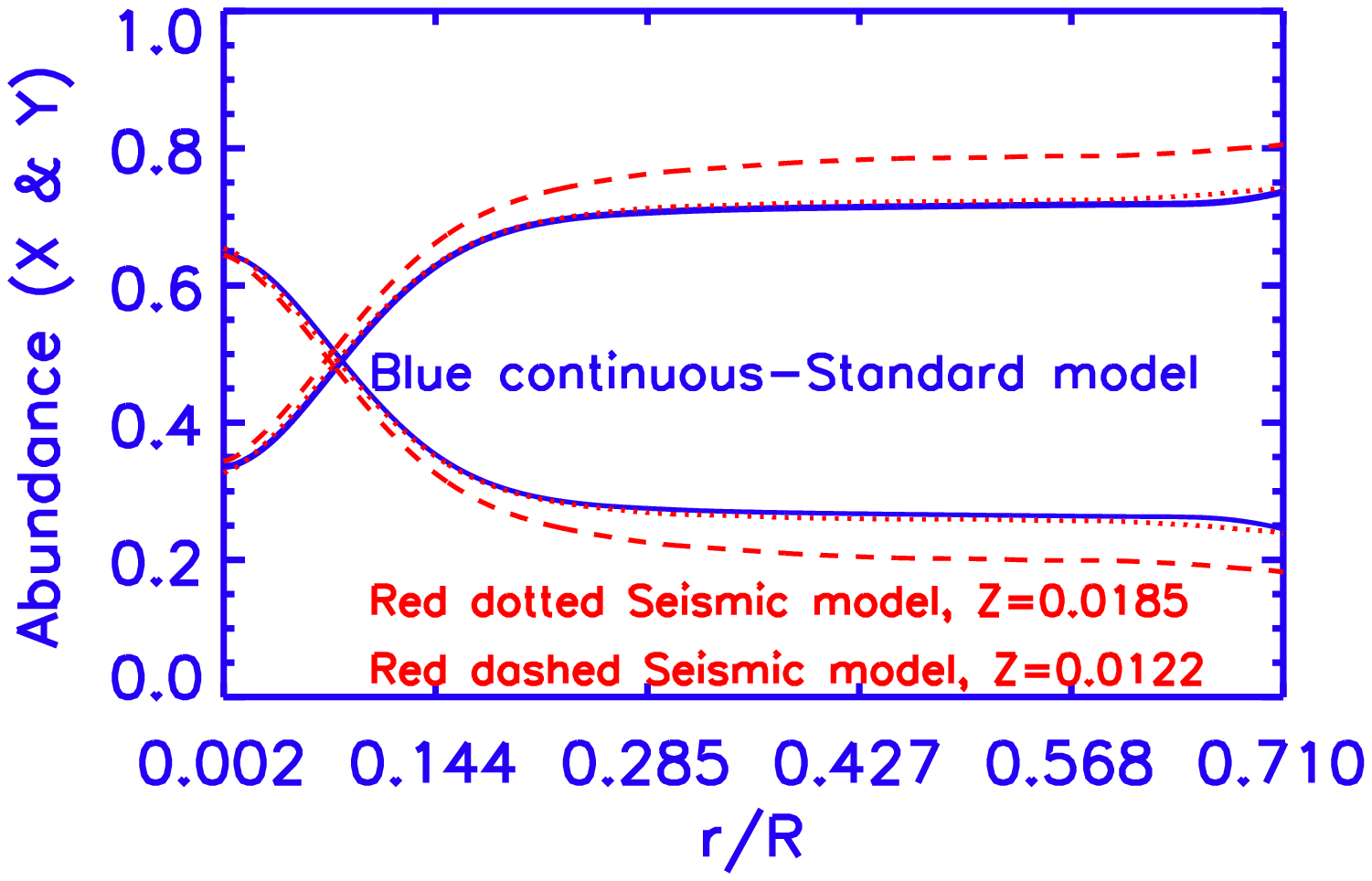}
}
\caption{Thermal structure such as pressure, temperature, density
and hydrogen abundance $X$ and helium abundance $Y$ obtained
by the solar seismic model for surface heavy elemental abundances
$Z=0.0185$ and $Z=0.0122$ respectively. Units are in cgs scale.}
\end{center}
\end{figure}

With appropriate
boundary conditions and assuming $Z$ as constant, 
earlier models (Shibahashi, Takata and Tanuma 1994; Shibahashi and Takata 1996;
Shibahashi and Takata 1997; 
Takata and Shibahashi 1998; Antia and Chitre 1999) solved
above equations by considering the radiative core only.
Although deduced profiles of thermal structure are
almost similar to the profiles of solar standard models, there
is no guarantee  that these deductions satisfy the
observed luminosity and the mass. By considering the
inverted sound speed (Takata and Shibahashi 1998)
for the whole region from center
to the surface, a global seismic model (Shibahashi, 
Hiremath and Takata 1998a; Shibahashi, Hiremath and Takata 1998b;
Shibahashi, Hiremath and Takata 1999) is developed 
that satisfies both the observed luminosity and the mass. 
It is found that seismic model satisfying one solar mass 
at the surface varies strongly on the nature of the 
sound speed profile near the center and a weak function 
of either depth of the convection zone or heavy elemental abundance Z/X.
It is also found that one solar mass at the surface is satisfied 
for the sound speed profile which deviates at the center 
by $\sim$ 0.22\% from the sound speed profile of solar model (Cristensen Dalsgaard 1996), 
if we adopt the value of nuclear cross section factor $S_{11}$
 value of 4.07 $\times 10^{-22}$ keV barns. 
The resulting chemical abundances at base of the
 convective envelope are obtained to be $X=0.755$, $Y=0.226$ and $Z=0.0185$
in case computed chemical abundance ratio $Z/X=0.0245$ matches
with the earlier (Grevesse and Noels 1993) chemical abundance ratio.
Neutrino fluxes of different reactions are computed
and are not different than the neutrino fluxes
computed from the standard solar model. 
\begin{center}
\begin{tabular}[t]{|l|c|c|c|c|c|}
\multicolumn{6}{c}{Table 1. Neutrino Fluxes (in the units
of $10^{10}cm^{-2}sec^{-1}$)}\\
\multicolumn{6}{c}{ estimated by the solar seismic model.}\\ \hline

$Source$ & SSM & Seism1  & Seism2 & Seism3& Seism4  \\ \hline
   & Z=0.0185  & Z=0.0185 & Z=0.0165 &Z=0.015 & Z=0.0122\\
   & Z/X=0.0247  & Z/X=0.0247 & Z/X=0.0215 &Z/X=0.0188 & Z/X=0.0158\\
   & BCZ=0.709  & BCZ=0.709 & BCZ=0.709 &BCZ=0.709 & BCZ=0.72\\ \hline
pp & 6.0 & 6.01 & 6.05  & 6.07 &  6.1 \\
pep & 0.014 & 0.015 & 0.014  & 0.015 &  0.015 \\
hep & 8$\times10^{-7}$ & 0.3$\times10^{-9}$ & 0.3$\times10^{-9}$&1.3$\times10^{-7}$&1.32$\times10^{-7}$ \\
${^7}$Be & 0.47 & 0.44 & 0.42  & 0.41 &  0.379 \\
${^8}$B &5.8$\times10^{-4}$ & 4.5$\times10^{-4}$ &4.03$\times10^{-4}$   &3.76$\times10^{-4}$ &3.18$\times10^{-4}$ \\
$^{13}$N & 0.06 &0.057 & 0.047  &  0.04 &  0.0287 \\
$^{15}O$ & 0.05 & 0.052 & 0.042  & 0.036& 0.025 \\
$^{17}F$ & $5.2\times10^{-4}$ & $4.02\times10^{-4}$ & 3.2$\times10^{-4}$ &2.7$\times10^{-4}$ &1.92$\times10^{-4}$\\
\hline
\end{tabular}
\end{center}
Recently, by considering the inverted sound speed kindly
provided by Antia, thermal structure and neutrino
fluxes are computed. One can notice from the Fig 8
that the deduced internal structural parameters such as mass,
luminosity, pressure, density and temperature are
almost similar to the structural parameters
obtained by the standard solar models. Neutrino
fluxes computed by using sound speed profile
computed from the standard model and the seismic
models with different $Z$ values are presented in
Table 1. In Table 1, nomenclatures for SSM, Seism and BCZ are
standard solar model of Dalsgaard (1996), seismic model
and base of the convection zone respectively.

One can notice that, for the earlier
(Grevesse and Noels 1993) chemical abundance ratio $Z/X=0.0247$,   
neutrino fluxes (Table 1) computed from the seismic model
is not different from the neutrino fluxes
computed from the solar standard solar model
that lead to a turning point with a strong conclusion
that deficiency of neutrinos emitted by the
sun lies in the neutrino physics ruling out the
astrophysical solutions. Even changing of 
uncertain physics of the interior, such as
opacity, equation of state, etc., or chemical composition
(especially the heavy elemental abundance $Z$)
could not alleviate the solar neutrino problem.

However and surprisingly, for the recently determined heavy
elemental abundance ($Z/X=0.016$; seismic model 4, Table 1), neutrino flux, especially
for ${^8}$B is substantially
reduced and is almost similar to the observed
neutrino fluxes, although helium abundance deduced
by this seismic model at the
base of the convective envelope is very low ($\sim$ 0.18)
and is inconsistent with other cosmic helium abundances ($\sim$ 0.23).
Recent determination (Asplund 2004) of heavy elemental
abundances (that substantially lowered the
abundances of carbon, nitrogen, oxygen and neon) has given a rebirth of astrophysical
solution to the solar neutrino problem (Turck-Chièze
{\em et.al.} 2010; Turck-Chièze and Couvidat 2011). 

\subsection{Inference of Dynamic Structure}
The term {\em dynamic structure} is mainly due
to rotation of the whole sun, although large-scale weak meridional
flow ($\sim$ few meters/sec from equator to both the northern
and southern poles) and strong convective flows
in the interior exist.
Observations show that, with a typical linear
velocity of $\sim$ 2 km/sec, sun rotates 
differentially, rotating faster at the equator
and slower at the poles. As mentioned in section 3.4, rotation of the sun
lifts the degeneracy of the oscillations 
that leads to frequency of oscillations depend upon azimuthal
order $m$ and, similar to Zeeman effect, split the frequencies
yielding a relation $\omega_{m} = \omega_{0} + \Omega m$
(where $\omega_{0}$ is frequency of the oscillations
in the rotating frame of reference and $\Omega$
is the angular velocity of the sun).
In this simple description, angular velocity
$\Omega$ is assumed to be of rigid body rotation
and hence is independent of radius and latitude respectively.
Observations show that sun is rotating differentially
and, by neglecting the effects due to centrifugal
force and the magnetic field structure, equation
(4) can be modified (Unno {\em et.al.} 1989; Dalsgaard 2002; 
Dalsgaard 2003; Thompson {\em et.al.} 2003) as follows
\begin{equation}
\delta \omega_{nlm} = \omega_{nlm} - \omega_{nl0} = 
 m \int_{0}^{R} \int_{0}^{\pi} K_{nlm}(r,\theta)\Omega(r,\theta)rdrd\theta \, ,
\end{equation}
\noindent where the kernels (Hansen et. al. 1977; Cuypers 1980;
Schou {\em et. al.} 1994) $ K_{nlm}(r,\theta)$ involve the eigen
functions of oscillations in the nonrotating frame. These
kernels depend only upon $m^{2}$ and hence the
rotational splittings ($\omega_{nlm} - \omega_{nl0}$)
are odd function of $m$. As the kernels are also
symmetrical about the equator, the rotational splittings
sense only the symmetrical component of $\Omega$.   
From the observed rotational splittings and
the computed kernels $ K_{nlm}(r,\theta)$ from the
adiabatic oscillations, one can invert the equation
(14) and hence rotation rate $\Omega(r,\theta)$
of the whole region of the solar interior is obtained. There
are different inversional techniques (Dalsgaard and Schou 1988; Schou et. al. 1994; 
Pijpers and Thompson 1992, 1994; Sekii 1996; Kosovichev {\em et. al.} 1996;
Korzennik {\em et.al.} 1996; Antia, Basu and Chitre 1998; Thompson {\em et. al.}
 2003 and references therein;
Howe 2009 and references there in; Antia and Basu 2010) and inferred rotation
rate of the solar interior is presented in Fig 9.
\begin{figure}
\begin{center}
\includegraphics[height=4.5cm,width=8.0cm]{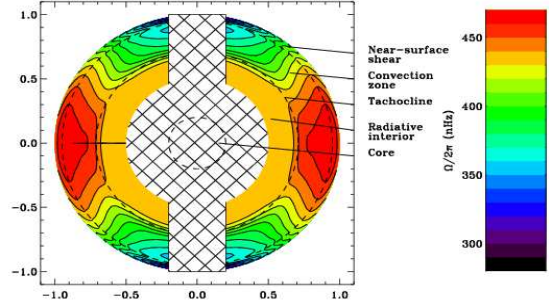}
\caption{ Isorotational contours inferred from 12 years average
data of MDI data. The cross hatched areas indicate
 the regions in which it is difficult or impossible to obtain reliable inversion results
with the available data (Courtesy : Howe).  } 
\end{center}
\end{figure}
One can notice from Fig 9 that sun rotates differentially
in the entire convective envelope and has a rigid body rotation
rate in the radiative core. In addition, near the surface
and the base of convection zone, two rotational shears
exist. The rotational shear near base of convection zone, 
where the transition from differential rotation to rigid
body rotation occurs, is called {\em tachocline} where
the seat of so called {\em dynamo} (that is believed to 
be origin and maintenance of solar cycle and activity
phenomena) is supposed to exist (Hiremath 2010; Hiremath and Lovely 2012
and references there in). However, yet it is not
completely understood how the sun attained such a rotational
structure in its interior. 
\subsection {Inference of Magnetic Field Structure}
Recent observations suggest that the entire universe
is pervaded by a large scale weak uniform magnetic field
structure (Ryu {\em et.al.} 2011 and references there in) and sun is also no exception. The sun is pervaded
by a large scale steady (diffusion time scale of  
$\sim$ billion yrs) global dipole like magnetic field structure with 
a  strength of $\sim$ 1 G (Stenflo 1993) and time dependent
magnetic field structure ($\sim$ kilo gauss)
that fluctuates over 22 years. If magnetic field
lines vary from pole to pole, such a geometry is called {\em poloidal
field} structure. Where as in case of {\em toroidal} field structure,
field lines are parallel to the equator. Sun is pervaded by a combined
weak ($\sim$ 1 G) poloidal and strong ($\sim$ $10^{3}$ G) toroidal
field structure. Sunspots are supposed to be formed due to
Alfven perturbations of the steady toroidal magnetic 
field structure (Hiremath and Lovely 2012) embedded in the solar interior. 
Large scale magnetic field structure induces
a force and distorts the geometrical figure of a star.
During the early evolutionary history of the sun, magnetic
field structure might have played a dominant role in
transferring angular momentum to the solar system.
This could be one of the main reasons why most of the 
angular momentum is concentrated in the solar system
rather than in the sun. It is believed that 22 yr magnetic cycle is due to
so called dynamo mechanism that is the result of interplay of
convection and rotation on the large scale poloidal magnetic field
structure although there are other alternative views (Hiremath 2010;
Hiremath and Lovely 2012). 

\subsubsection{Inference of Magnetic Field Structure : Primary Inversions}
Signatures of both the poloidal and toroidal magnetic
field structures can be found in the observed even degree
frequency splittings. In addition, second order rotational
effect (where as first order or linear rotational effect
is sensed by the odd degree frequency splittings as
given in section 4.2) and aspherical sound speed perturbations
(Kuhn 1988) also contribute to the even degree frequency
splittings. Hence, by assuming that aspherical sound speed
perturbations are negligible, with the internal rotational
structure as inferred from the helioseismology (section 4.2),
contribution due to second order rotational effects is computed and is subtracted
from the even degree frequency splittings and, 
resulting residual of even degree splittings are 
compared with the computed frequency splittings due
to assumed magnetic field structure.
Detailed inversion procedures can be found in the previous
studies (Dziembowski and Goode 1989; Gough and Thompson
1991; Dziembowski and Goode 1991; Basu 1997; Antia, Chitre and Thompson 2000;
Antia 2002; Baldner {\em et. al.} 2010).
Recently Baldner {\em et. al.} (2010) came to
the conclusion that the observed variation of even degree frequency splittings can be 
explained if the sun is pervaded by a right combination
of poloidal ($\sim$ 100 G)  and toroidal (that varies from $\sim$ $10^{3}$ G near the surface
to $10^{4}$ G near base of the convection zone) magnetic field structure. 
By considering the SOHO/MDI magnetograms and from the analysis of sunspot data
during their initial appearance on the surface,
recently, Hiremath and Lovely (2012) estimated strength of toroidal
magnetic field structure of similar order confirming
the previous theoretical estimates (Choudhuri and Gilman 1987;  D'Silva and Choudhuri 1993;
D'Silva and Howard 1994; Hiremath 2001). 
 \subsubsection{Inversion of Poloidal Magnetic Field Structure : Secondary Inversions} 
Owing to large diffusion time scales due to large dimension and finite conductivity
of the solar plasma, sun might have retained a large scale poloidal magnetic
structure from its protostar phase even after the
convective Hayashi phase (Cowling 1953; Layzer, Rosner and
Doyle 1979; Piddington 1983; Mestel and Weiss 1987; Spruit 1990).
Presence of such a large scale poloidal magnetic
field structure, especially during the solar minimum,
 can be clearly discerned during the solar total eclipse.
Field lines of dipole like magnetic field structure are
clearly delineated in the intensity patterns of the white light
pictures taken during the eclipse. 
If the Lorentz force due to either poloidal
or toroidal magnetic field structure is
very weak compared to dynamical effects such
as rotation, one can show (Hiremath 1994; Hiremath and
Gokhale 1995) that weak poloidal magnetic field structure isorotates
with the solar plasma. In fact, in case of the sun, especially
for the poloidal magnetic field structure ($\sim$ 1 G),
this condition is valid. That means, if $\Omega$ is
angular velocity of the plasma and $\Phi$ is the
magnetic flux function representing flux of
the poloidal magnetic field structure, one
can show that 
\begin{equation}
\Omega = function (\Phi)  \, .
\end{equation} 
If one knows the internal rotational
structure of the sun, by suitable combination of poloidal
magnetic field structure that is obtained consistently
from the solution of MHD equations, one can satisfy the above
condition. To a first approximation, right hand side of equation
can be linearized of the form

\begin{equation}
\Omega = \Omega_{0} + \Omega_{1} (\Phi) \, .
\end{equation}  
\begin{figure}
\begin{center}
\includegraphics[height=6.0cm,width=6.0cm]{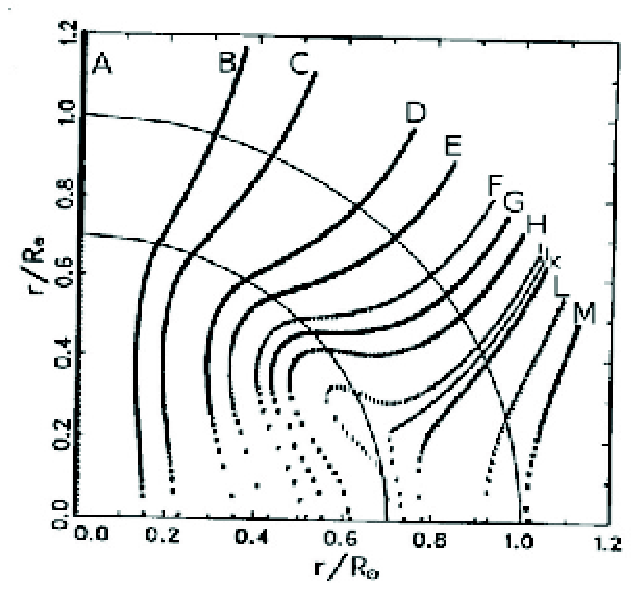}
\includegraphics[height=6.0cm,width=6.0cm]{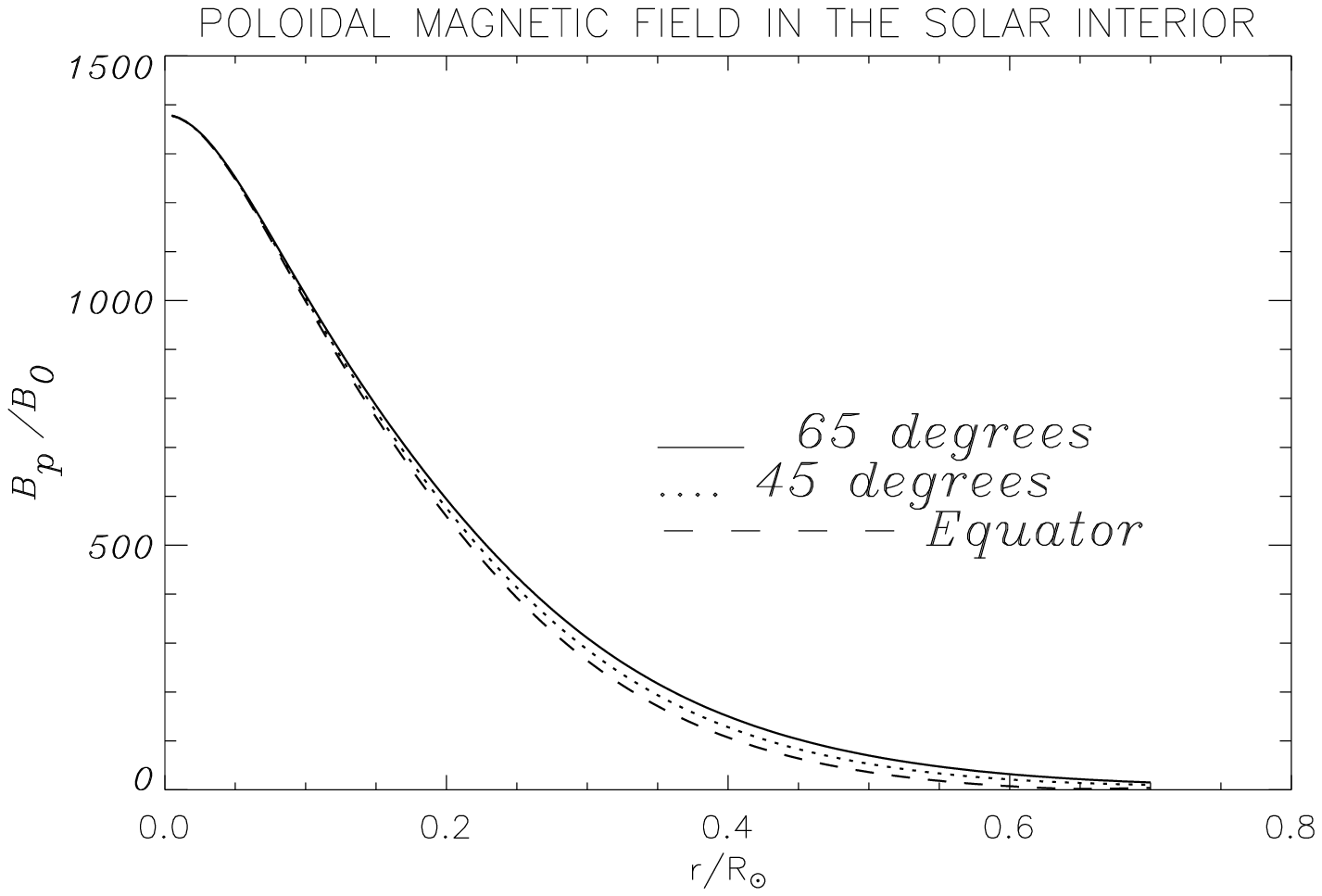}
\caption{Left figure illustrates the meridional cross
section of poloidal component of magnetic field structure.
This figure is reproduced from, and all rights are reserved by, 
the Astrophysical Journal. Where as right figure illustrates
radial and latitudinal variations of the magnitude
of the poloidal magnetic field structure $B_{p}$ normalized
to an asymptotic uniform magnetic field that merges
with the interplanetary field structure.} 
\end{center}
\end{figure}

With reasonable assumptions and approximations and by
using Chandrasekhar's (1956a; 1956b) MHD equations
for the case of axisymmetry and incompressibility,
Hiremath and Gokhale (1995) have shown that
poloidal component of the sun's steady magnetic field structure can be
modeled as an analytical solution of magnetic diffusion
equation and the magnetic flux function $(\Phi)$
for both the radiative core (RC) and the convective
envelope (CE) are expressed as
\begin{equation}
\Phi_{RC}(x, \vartheta) = 2 \pi A_{0}R_{c}^{2} x^{1/2} sin^{2} \vartheta 
\sum_{n=0}^{\infty} \lambda_{n} J_{n+3/2}(\alpha_{n}x)C_{n}^{3/2}(\mu) \, ,
\end{equation} 
where $x=r/R_{c}$, $R_{c}$ is radius of the radiative
core, $\mu=cos \vartheta$, $r$ and $\vartheta$
are radial and colatitude coordinates, $n$ is non negative
integer, $C_{n}^{3/2}$ is the Gegenbaur's polynomial 
of degree $n$, $ J_{n+3/2}(\alpha_{n}x)$ is Bessel
function of order $(n+3/2)$ and $\alpha_{n}$ are
the eigen values that are to be determined from the
boundary conditions. Here $A_{0}$ is taken as a scale factor for the field
and hence $\lambda_{n}= A_{n}/A_{0}$. Similarly, magnetic flux
function for the case of convective envelope is
\begin{equation}
\Phi_{CE}(x, \vartheta) = \pi B_{0} R_{c}^{2} sin^{2} \vartheta
 \big [ x^{2}C_{0}^{3/2}(\mu) + \sum_{n \ge 0 }\hat \mu_{n+1} x^{-(n+1)} C_{n}^{3/2} (\mu) \big ] \, ,
\end{equation}
where $\hat \mu_{n+1} = M_{n}/(\pi B_{0} R_{c}^{n+3})$ are 
strengths of multipoles that are scaled to 
an asymptotically uniform magnetic field structure $B_{0}$. One can notice
from the above equation that magnetic field structure approaches
an asymptotically uniform field structure as $x \rightarrow \infty$.   
With appropriate boundary conditions at base of
the convection zone and with helioseismically inferred rotation, 
one can uniquely estimate the unknown coefficients
and eigen values respectively. In case of sun and with the
available helioseismically inferred rotation (Dziembowski {\em et. al.} 1989;
Antia, Basu and Chitre 1998),
geometry of the poloidal field structure is computed (Hiremath 1994; Hiremath and Gokhale 
1995) and is presented in Fig 10. If one considers strength of interplanetary magnetic field structure
($\sim 10^{-5}$ G) as a scaling factor, strength of sun's poloidal magnetic field structure 
near its center is found to be $\sim 10^{7}-10^{8}$ G. From the dynamical
constraints, one can rule out such a strong magnetic field structure
near the center. However, if the 22 yr solar cycle and activity
phenomenon is due to global slow MHD
modes (Hiremath and Gokhale 1995), $B_{0}$ is estimated
to be $\sim 10^{-2}$ G and, in that case strength of poloidal 
magnetic field structure near the center turns out be $\sim 10^{5}$ G. Off course
one would have found signature of such a strong field structure in
 the observations of $p$ mode oscillations. Unfortunately low degree
`p' modes that penetrate deeply in the radiative core can not sense
close to the center. Hence, one has to wait for the discovery
of elusive $g$ modes that are highly sensitive to the central structure
of the sun and hence their characteristics are modified (Rashba 2006, 2007, 2008;
Burgess {\em et. al.} 2004) by such a strong magnetic field structure.   
\section {Concluding Remarks}
Although seismology of the sun has provided some of the detailed
global aspects of thermal and dynamic structures 
of the sun's interior, other global structures such as detailed 
informations regarding meridional
flow (this information is needed to accept or reject so called flux transport
dynamo models) and magnetic field structure are necessary. Hitherto solar seismologists
came to the conclusion that the difference between the sun's
thermal structure (especially the sound speed) and the modeled
thermal structure is almost similar. However, recent revision
of heavy-element abundances (Asplund {\em et. ali.} 2000, 2004;
 Allende Prieto {\em et.al.} 2001, 2002; Asplund {\em et. ali.} 2005) that are reduced
drastically, further worsened the difference between the sun's
and modeled thermal structure leading to revisit of 
exotic astrophysical solutions that are used to explain the solar 
neutrino problem. Importantly, as noticed by the
previous studies (Chaplin and Basu 2008; Basu and Antia 2008),
critical examination of modeling and method
of computation in estimating the heavy-elemental
abundance are necessary.

As for the dynamical helioseismic inferences, especially the rotational structure,
`p' modes are not the true representatives to sense the deep radiative core.
For this purpose one has to wait for discovery of `g' modes
that have high sensitivity near the core.
Hence knowing of magnitude and form ({\em i.e.,}
dependent or independent of radius and
latitude) of thermal and rotational structures near
the core are essential for understanding overall structure
and temporal variations  of the sun on short ($\sim$ 11 years) 
and long ($\sim$ billion years) time scales.
Although helioseismic inferences of rotation rate of the
convective envelope ruled out the increase of rotation rate
(as is assumed to be in the earlier turbulent dynamo models) 
from surface to the interior and cylindrical
isorotational contours (simulations by Gilman and Miller 1981), physics
of isorotational contours, especially near the surface, is completely not understood.
However, it should be appreciated that the simulated isorotational
contours from base of convection zone to $0.935R_{\odot}$
(Kitchatinov and Rudiger 1995; Elliot {\em et. al} 2000;
Robinson and Chan 2001, Hiremath 2001; Brun and Toomre 2002, 
Rempel 2005; Meisch {\em et. al.} 2006; Kuker, Rudiger and Kitchatinov 2011;
Brun {\em et.al} 2011)
are almost similar to the rotational isocontours as inferred
from helioseismology although none of the present simulations 
reproduce the near surface decreasing rotational profile
 from $0.935R_{\odot}$ to the surface. Even if numerical
simulations achieve the goal of reproducing near
surface rotational profile, from the MHD stability 
criterion (Dubrulle and Knobloch 1993; Mestel 1999),
$r^{2}{d \over dr}[{\Omega^{2}\over r^{2}}] - ({1\over r ^{2}}) {d\over d r}[{(r^{2}B_{\phi}^{2}) \over 4\pi\rho}]>0 $ (where $r$ is the radial coordinate,
$\Omega$ is angular velocity, $\rho$ is density and
$B_{\phi}$ is toroidal magnetic field structure), one can show 
that such a decreasing rotational profile near the surface is 
unstable (independently, Miesch and Hindman 2011, recently came
to a similar conclusion) unless sun acquired near equipartition
toroidal magnetic field structure. Interestingly,
toroidal field structure estimated by theoretical 
(Hiremath 2001; Brun, Miesch and Toomre 2004) and helioseismic
(Antia 2002; Baldner {\em et.al.} 2010) inferences yield of 
similar magnitudes. Hence, near surface rotational profile is
stable. 

However, why the sun acquired such a decreasing near surface
rotational profile and hence the toroidal magnetic field structure
are remained to be explained. Although beyond the scope of this
presentation, such a near surface decreasing 
rotational profile can be explained if one speculates that during the early history
of the planetary formation, protoplanetary mass that rotated
with Keplerian velocity might have accreted on to the sun
that resulted in winding of threaded ambient large scale poloidal magnetic field structure
 into toroidal magnetic field structure. If this speculation
is correct, such a protoplanetary accretion scenario might give
some clues regarding absence of super earths near vicinity of
the sun and peculiar high solar mass (Gustafsson 2008) compared to other stars
in the sun's neighborhood in the Galaxy.  
Incase we accept the accretion scenario (during
the early solar system formation), longstanding
problems such as angular momentum and planetary formation
of the solar system can be solved. With such an accretion scenario
and hence high solar mass, unsolved faint young sun paradox (Minton  and Malhotra 2007) and recent
puzzle of computation of low heavy elemental abundances
(Asplund {\em et. al.} 2005) that are incompatible (Basu {\em et al.} 2007;
Basu and Antia 2008) with the helioseismic inferences
can also be alleviated upto some extent (Winnick {\em et. al.} 2002; Haxton  and Serenelli 2008;
Nordlund 2009; Meléndez {\em et. al.} 2009; Guzik and Mussack 2010 and references
there in; Serenelli {\em et.al.} 2011).

As for the secondary inversion of poloidal magnetic field
structure, this method requires the rotational profile
of the interior as inferred by the helioseismology
for estimation of different magnetic moments.
Instead, as the weak ($\sim$ 1 G) poloidal magnetic field
structure isorotates with the internal rotation of
the plasma and hence rotation is a function of magnetic
flux, equation (14) can be used for simultaneously
inverting rotational and poloidal magnetic field
structures of the interior (Gokhale and Cristensen Dalsgaard, 
private communication).

\centerline{\bf References}

\noindent Agnihotri, R., Dutta, K and Soon, W. 2011, JASTP, 73, 1980
 
\noindent Anguera Gubau, M., Palle, P. L., Hernandez, P and Roca-Cortes, T. 1990, Sol Phys, 128, 79

\noindent Antia, H. M., Basu, Sarbani \& Chitre, S. M. 1998, MNRAS, 298, 543

\noindent Antia, H. M and Chitre, S. M. 1999, A\&A, 347, 1000

\noindent Antia, H. M., Chitre, S. M., \& Thompson, M. J. 2000, A\&A, 360, 335

\noindent Antia, H. M. 2002, Proceedings of IAU Coll. 188, ESA SP-505, p.71

\noindent Antia, H. M. \& Basu, S., 2010, ApJ, 720, 494

\noindent Allende Prieto, C., Lambert, D.L., Asplund, M. 2001. ApJ. 556, L63

\noindent Allende Prieto, C., Lambert, D.L., Asplund, M. 2002. ApJ. 573, L137

\noindent Asplund, M., Nordlund, Trampedach, R., Stein, R.F. 2000, A \& A, 359, 743

\noindent Asplund, M., Grevesse, N., Sauval, A.J., Allende Prieto, C., Kiselman,
D. 2004, Astron. Astrophys. 417, 751 

\noindent Asplund, M., Grevesse, N and  Sauval, A. J. 2005, ASP Conf. Series 336, p. 25

\noindent Baldner, C. S., Antia, H. M; Basu, S and Larson, T. P. 2010, Astronomische Nachrichten, 331, 879

\noindent Basu, S. 1997, MNRAS, 288, 572

\noindent Basu, S., Chaplin, W. J., Elsworth, Y., New, R., Serenelli, A. M., \& Verner, G. A. 2007, ApJ, 655, 660

\noindent Basu, S and  Antia, H. M. 2008, Physics Reports, 457, 217

\noindent Basu, S., Chaplin, W. J., Elsworth, Y, New, R and Serenelli, A. M. 2009, ApJ, 699, 1403
 
\noindent Basu, S. 2010, Astrophys \& Space Sci, 328, 43

\noindent Brun, A. S., Miesch, M. S and Toomre, J. 2004, ApJ, 614, 1073

\noindent Brun, A. S. \& Toomre, J. 2002, ApJ, 570, 865 

\noindent Brun, A. S., Miesch, M. S., \& Toomre, J. 2011, ApJ, 742, 79

\noindent Burgess, C. P., Dzhalilov, N. S., Rashba, T. I.,                      
Semikoz, V. B and Valle, J. W. F., 2004, MNRAS, 348, 609

\noindent Chandrasekhar, S. 1956a, ApJ, 124, 232

\noindent Chandrasekhar, S. 1956b, ApJ, 124, 244

\noindent Chandrasekhar, S. 1964, ApJ, 139, 664

\noindent Chaplin, W. J and Basu, S, 2008, Sol Phys, 251, 53

\noindent Choudhuri, A. R and Gilman, P. A. 1987, ApJ, 316, 788

\noindent Cowling, T. G. 1953, MNRAS, 94, 39

\noindent Cuypers, J. 1980, A\&A, 89, 207

\noindent Dalsgaard, C., Gough, D. O and and Libbrecht, K. B. ApJ, 341, L103

\noindent Dalsgaard, C., et al. 1996, Science, 272, 1286 

\noindent Dalsgaard, C. 2003, Lecture Notes on Stellar Oscillations, p. 66

\noindent Dalsgaard, C. 2002, Rev. Mod. Phys. 74, 1073

\noindent Deubner, F.-L. 1975, A\&A, 44, 371

\noindent Deubner, F.-L and Gough, D. 1984, ARA \& A, 22, 593

\noindent D'Silva, S., \& Choudhuri, A. R. 1993, A\&A, 272, 621

\noindent D'Silva, S., \& Howard, R. 1994, Sol Phys, 151, 213

\noindent Dubrulle, B \& Knobloch, E.,  1993, A\&A, 274, 667

\noindent Dziembowski, W. A and Goode, P. R. 1989, ApJ, 347, 540

\noindent Dziembowski, W. A., Goode, P. R and Libbrecht, K. G. 1989, ApJ, 343, L53

\noindent Dziembowski, W. A and Goode, P. R. 1991, ApJ, 376, 782

\noindent Elliott, J. R., Miesch, M. S., \& Toomre, J. 2000, ApJ, 533, 546 

\noindent Feynman, J. 2007, AdSpR, 40. 1173

\noindent Gilman, P. A and Miller, J. 1981, ApJS, 46, 211

\noindent Goldreich, P and Kumar, P. 1988, ApJ, 326, 462

\noindent Gough, D. O and Taylor, P. P. 1984, Mem. Soc. Astron. Ital., 55, 215

\noindent Gough, D. O and Thompson, M. J. 1991, in Solar Interior and Atmosphere, . p. 519

\noindent Grec, G., Fossat, E and Pomerantz, M. 1983, Sol Phys, 82

\noindent Grevesse, S and Noels, A. 1993, : in Origin and Evolution of the Elements, p. 15

\noindent Guzik, J. A and Mussack, K., 2010, ApJ,  713, 1108

\noindent Haigh, J. D. 2007, Living Rev Sol Phys, 4, 2

\noindent Hansen, C. J., Cox, J. P and Van Horn, H. M. 1977, ApJ, 217, 151

\noindent Haxton, W. C. and Serenelli, A. M. 2008, ApJ, 687, 678

\noindent Hill, G., Deubner, F. L and Isaak, G. 1991, in Solar Interior and Atmosphere,
p. 329

\noindent Hiremath, K. M. 1994, PhD Thesis, Bangalore University

\noindent Hiremath, K. M and Gokhale, M. H. 1995, ApJ, 448, 437

\noindent Hiremath, K. M. 2001, Bull Astron Soc India, 29, 169
 
\noindent Hiremath, K. M and Mandi, P. I. 2004, NewA, 9, 651

\noindent Hiremath, K. M. 2006a, Journal of Astrophysics and Astronomy, vol 27, 367

\noindent Hiremath, K. M. 2006b, Journal of Astrophysics and Astronomy, vol 27, 277

\noindent Hiremath, K. M. 2006c, in the proceedings of ILWS workshop, Goa, p. 178

\noindent Hiremath, K. M. 2008, Ap \& SS, 314, 45
 
\noindent Hiremath, K. M. 2009, Sun \& Geosphere, 4, 16  

\noindent Hiremath, K. M. 2010, Sun \& Geosphere, 5, 17

\noindent Hiremath, K. M and Lovely, M. R. 2012, New Astronomy, 17, 392

\noindent Howe, R. 2009, Living Review in Solar Phys, 6, 1

\noindent Kitchatinov, L. L. and Rüdiger, G. 1995, A\&A, 299, 446 

\noindent Kuhn, J. R. 1988, ApJ, 331, L131

\noindent Kuker, M., Rudiger, G and Kitchatinov, L. L. 2011, A\&A, 530, 48

\noindent Kumar, P., Franklin, J and Goldreich, P. 1988, ApJ, 328, 879

\noindent Kumar, P and Goldreich, P. 1989, ApJ, 343, 558

\noindent Korzennik, S. G., Thompson, M. J and Toomre, J. 1996, in IAU Symp 181, p. 211

\noindent Kotov, V. A., Severnyi, A. B and  Tsap, T. T. 1978, MNRAS, 183, 61

\noindent Layzer, D., Rosner, R and Doyle, H. T. 1979, ApJ, 229, 1126

\noindent Leibacher, J. W and Stein, R. F. 1971, ApJL, 7, 191L

\noindent Leighton, R. B. 1960, IAUS, 12, 321

\noindent Leighton, R. B., Noyes, Robert W and Simon, George W. 1962, ApJ, 135, 474

\noindent Libbrecht, K. G. 1988, ApJ, 334, 510

\noindent Libbrecht, K. G. 1989, ApJ, 336, 1092

\noindent Lynden-Bell, D and Ostriker, J. P. 1967, MNRAS, 136, 293

\noindent Miesch, M. S., Brun, A. S and Toomre, J. 2006, ApJ, 641, 618

\noindent Miesch, M. S and Hindman, B. W. 2011, ApJ, 743, 79

\noindent  Minton, D. A  and  Malhotra, R., 2007, ApJ, 660, 1700

\noindent Meléndez, J., Asplund, M., Gustafsson, B and Yong, D. 2009, ApJ, 704, L66

\noindent Mestel, L and Weiss, N. O. 1987, MNRAS, 226, 123

\noindent Musman, S and Rust, D. M. 1970, Solar Phys, 13, 261

\noindent Nishikawa, J., Hamana, S., Mizugaki, K and Hirayama, T. 1986, 
Publ. Astron. Soc. Japan, 38, 277

\noindent Nordlund, A. 2009, arXiv:0908.3479

\noindent Palle, P. L., Regulo, C and Roca-Cortes, T. 1989, A \& A, 224, 253

\noindent Perry, C, A. 2007, AdSpR, 40, 353

\noindent Piddington, J. H. 1983, Astrophys Space Sci, 90, 217

\noindent Pijpers, F. P and Thompson, M. J. 1992, A\&A, 262, 33

\noindent Pijpers, F. P and Thompson, M. J. 1994, A\&A, 281, 231

\noindent Rashba, T. I., Semikoz, V. B and Valle, J. W. F., MNRAS, 370, 845, 2006

\noindent Rashba, T. 2008, Journ Physics: Conf Ser, 118, p. 012085

\noindent Rashba, T. I., Semikoz, V. B., Turck-Chièze, S and Valle, J. W. F. 2007, MNRAS, 377, 453 

\noindent Reid, G. C. 1999, JASTP, 61, 3

\noindent Rempel, M. 2005, ApJ, 622, 1320 

\noindent Robinson, F. J and Chan, K. L. 2001, MNRAS, 321, 723

\noindent Ryu, D., Schleicher, D. R. G. Treumann, R. A., Tsagas, C. G., Widrow, L. M.
2011, Space Science Reviews, 312

\noindent Scafetta, N and West, B. J., 2008, Physics Today, March Issue, p. 50-51

\noindent Scherrer, P. H., Wilcox, J. M., Kotov, V. A., Severny, A. B and Tsap, T. T. 1979
Nature, 277, 635

\noindent Scherrer, P. H and Wilcox, J. M. 1983, Sol Phy, 82, 37

\noindent Schou, J., Christensen-Dalsgaard, J and  Thompson, M. J. 1994, ApJ, 433, 389

\noindent Serenelli, A. M., Haxton, W. C \& Pena-Garay, C., 2011, arXiv:1104.1639

\noindent Sekii, T. 1996, in IAU Symp 181, p. 189

\noindent Severnyi, A. B., Kotov, V. A.,  Tsap, T. T. 1976, Nature, 259, 87

\noindent Severnyi, A. B., Kotov, V. A and Tsap, T. T. 1979, Soviet Astronomy, 23 641

\noindent Shibahashi, H., Takata, M and Tanuma, S. 1995, ESA SP, Proceedings of the 4th Soho Workshop, p. 9

\noindent Shibahashi, H and Takata, M. 1996, PASJ, 48, 377

\noindent Shibahashi, H and Takata, M. 1997, 1997, IAU Symp 181, 167

\noindent Shibahashi, H., Hiremath, K. M and Takata, M. 1998a, ESASP, 418, 537

\noindent Shibahashi, H., Hiremath, K. M and Takata, M. 1998b, IAU Symp, 185, 81

\noindent Shibahashi, H., Hiremath, K. M and Takata, M. 1999, Advance Space Res, 24, 177

\noindent Shine, K, P. 2000, SSRv, 94, 363

\noindent Soon, W. 2005, Geophys. Res. Let, 32, 16, L16712

\noindent Spruit, H. C. 1990, in Inside the Sun, p. 415

\noindent Stenflo, J. O. 1993, in Solar Surface Magnetism

\noindent Takata, M and Shibahashi, H. 1998, ApJ, 504, 1035 

\noindent Thompson, Michael J., Christensen-Dalsgaard, Jørgen, Miesch, Mark S. \& Toomre, J. 2003, 
Astron Astrophys Rev, 41, 599

\noindent Tiwari, M and Ramesh, R. 2007, Current Science, vol 93, 477

\noindent Turck-Chièze, S., Palacios, A., Marques, J. P and Nghiem, P. A. P. 2010, ApJ, 715, 1539

\noindent Turck-Chièze, S and Couvidat, S. 2011, Rep Prog Phys, 74, 086901

\noindent Ulrich, Roger K. 1970, ApJ, 162, 99

\noindent Unno, W., Osaki, Y., Ando, H., Sao, H and Shibahashi, H. 1989, 
in "Noradial Oscillations of Stars", p. 108

\noindent Winnick, R. A., Demarque, P., Basu, S., \& Guenther, D. B. 2002, ApJ, 576, 1075
\end{document}